\documentclass[superscriptaddress,twocolumn,prx,aps,preprintnumbers,notitlepage,longbibliography]{revtex4-1}
\usepackage[latin9]{inputenc}
\usepackage{amsmath}
\usepackage{amssymb}
\usepackage{graphicx}
\usepackage{esint}
\usepackage[scr=boondoxo, scrscaled=1.05]{mathalfa}
\allowdisplaybreaks

\makeatletter

\pdfpageheight\paperheight
\pdfpagewidth\paperwidth

\pdfoutput=1

\usepackage{epsfig}
\usepackage{graphics}

\usepackage{mciteplus}
\mciteErrorOnUnknownfalse

\makeatother

\begin{document}

\preprint{FERMILAB-PUB-19-334-A}

\title{Pattern of   perturbations from a coherent quantum inflationary horizon}

\author{Craig  Hogan}
\affiliation{University of Chicago, 5640 South Ellis Ave., Chicago, IL 60637}
\affiliation{Fermi National Acclerator Laboratory, Batavia, IL 60510}
\begin{abstract}
It is proposed that if  quantum states of space-time are coherent on null surfaces,  holographic Planck-scale  fluctuations of  inflationary horizons dominate the formation of  primordial scalar curvature perturbations. It is shown that the reduction of quantum states on nearly-spherical emergent horizon surfaces around each observer creates  a distinctive pattern whose correlations  in the angular domain differ from the standard quantum theory of inflation. Causal constraints are used in a semiclassical model to  formulate candidate directional symmetries.  It is suggested that this hypothesis could provide a  physical explanation for several well known anomalies measured  in CMB anisotropy.  
New exact symmetries are predicted, such as a vanishing temperature correlation function at 90 degrees angular separation, that can be tested with current data.
\end{abstract}

\maketitle

\section{Introduction}

A standard cosmological model\cite{Weinberg:2008zzc}  is now supported   
by  a considerable body of evidence, especially precise measurements of  correlations in cosmic microwave background radiation (CMB)\cite{2013ApJS..208...20B,2013ApJS..208...19H,WMAPanomalies,Ade:2015xua,Ade:2015lrj,Array:2015xqh,Ade:2015hxq,Aylor:2017haa,Akrami:2018vks,Aghanim:2018eyx,Akrami:2018odb,Akrami:2019bkn}.  The early evolution is  generally described by slow-roll inflation\cite{Kadota:2005hv,Baumann:2009ds,Kamionkowski:2015yta}, during which the repulsive gravity of an exotic, metastable scalar inflaton field vacuum drives an accelerating expansion. 
Inflation shapes the structure of the universe on the largest scales--- a large, nearly-uniform, nearly-flat  geometry, with nearly-scale-invariant primordial perturbations in curvature that give rise to cosmic structure.
In the standard picture, the cosmic perturbations are created by  quantum fluctuations of  the  inflaton field vacuum,  coupled by linearized gravity to the classical background geometry. 

It is  possible that  all the classical elements of this picture are essentially correct, but that the quantum mechanical model is radically incomplete, because its approximations omit a fundamental coherence of emergent quantum gravity on causal surfaces.  Instead of linearized quantum fields, 
cosmic perturbations  could be dominated by spatially-coherent holographic degrees of freedom of Planck scale quantum geometry.  Such holographic or ``spooky'' scenarios\cite{Banks:2018ypk,PhysRevD.99.063531} do not assume a pre-existing, determinate classical background metric; instead,
the  perturbations are associated with the noisy emergence of space and time from a  quantum system. Some theoretical motivations for this approach are  summarized in the Appendix below.


Unlike the standard  scenario based on quantized inflaton field modes, holographic inflation posits that  the inflationary horizon of every observer, defined as its past light cone at the end of inflation, is a  coherent nonlocal quantum object, like a whole atom. A  similar  hypothesis has recently been applied to 
the horizon in some quantum models of  black holes
\cite{Hooft:2016cpw,Hooft:2016itl,Hooft2018,Giddings:2018koz,Giddings:2019vvj}.
In this picture, the quantum state of the horizon is a superposition of slightly deformed horizon surfaces, whose deformations are coherent at large angular separations.
The cosmic time at which decoherence occurs at a given comoving position depends on the observer.
Spatial projections and correlations of geometrical quantum states ``collapse'' onto the  nearly-spherical  horizons instead of spatially infinite plane wave modes,
so that  primordial  perturbations  are laid down coherently,  with nonlocal quantum correlations, on a nearly-spherical horizon in all directions. 

The magnitude of the holographic perturbations depends only on $H$,  the expansion rate during inflation in units of the  Planck time $t_P\equiv \sqrt{\hbar G/ c^5}$. 
Because of the coherence,  the dimensionless perturbation power of scalar perturbations $\langle{\Delta}^2\rangle\approx Ht_P$
from Planck scale quantum gravity is much larger than the standard perturbations from quantum perturbations of inflaton field modes on scale, $\langle{\Delta}^2\rangle\approx (Ht_P)^2 \epsilon^{-1}$, where $\epsilon$ denotes a slow-roll parameter of the inflaton potential. Inflation therefore occurs at a  lower $H$ than in typical standard models, but because it also depends on a slowly varying $H$, 
the predicted primordial power spectrum of curvature in the holographic picture  is indistinguishable from standard quantum inflation\cite{PhysRevD.99.063531}: holographic inflation still  produces  perturbations with a nearly scale invariant,  slightly tilted power spectrum.
Post-inflation evolution is  standard,  so it preserves the precise match of standard cosmology to a host of measurements that depend only on the power spectrum, including  CMB anisotropy spectra and  measurements of cosmic large structure over a large range of scales.

The most distinctive observable relics of  a coherent horizon\cite{PhysRevD.99.063531} are new
 correlations in the initial phase and direction of classical curvature perturbations. 
 It is proposed here that  directly measurable signatures of these holographic correlations  may appear in specific properties  of the  pattern of  CMB  anisotropy in the angular domain. They  take the form of precisely defined symmetries or constraints that generally do not occur in any particular realization in the standard picture,   because of cosmic variance. 
 Simple constraints from causal structure and rotational symmetry in the emergent system are used here to derive precisely defined
candidate  symmetries of correlations at large angular  separations.  

   The properties of the pattern derived here are better motivated, and more concretely specified, than previously  conjectured holographic correlations\cite{Hogan:2002xs,Hogan:2003mq,10.2307/27857721}.
They could provide a unified physical  explanation of some long-known, seemingly unrelated statistical  anomalies in the CMB\cite{WMAPanomalies,Ade:2015hxq,Akrami:2019bkn}.   Moreover,  they lead to new predictions  for properties that have no particular significance in the standard picture: for example, the most robust new prediction here is  that the angular  correlation function of curvature should exactly vanish at 90 degree angular separation.  Perturbations can also naturally be generated with significant  anticorrelation at large angular separations, at a level unlikely to occur in the standard picture. These  predictions can be used  to implement new,  sharply formulated statistical comparisons with the standard quantum model of  inflation.
 
\section{Holographic perturbations}

\begin{figure}
\begin{centering}
\includegraphics[width=\linewidth]{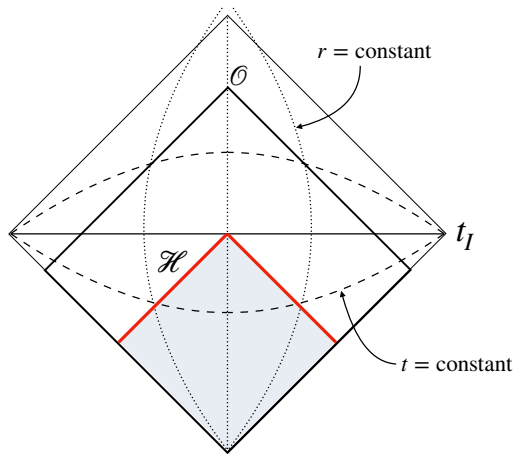}
\par\end{centering}
\protect\caption{Causal structure of a classical inflationary universe, with two spatial dimensions suppressed. Vertical axis represents the  world line of an observer $\cal O$, horizontal axis represents 
the end of inflation $t_I$, and left and right halves represent antipodal spatial directions. The horizon $\cal H$ forms the outer boundary of   causal diamonds that end on $\cal O$ before $t_I$.  Dashed lines represent spatial hypersurfaces of constant cosmic time $t$,  and dotted lines are world lines of constant comoving position $r$. In standard inflation,  the amplitude of a plane wave mode freezes out acausally everywhere at the time  when its wavelength matches the horizon, so quantum states of geometry collapse into eigenstates of wave modes on 3D spacelike hypersurfaces of constant  $t$; in holographic inflation, geometrical states collapse on   boundaries of causal diamonds defined by $\cal H$.
  \label{penrose}}
\end{figure}

\begin{figure}
\begin{centering}
\includegraphics[width=\linewidth]{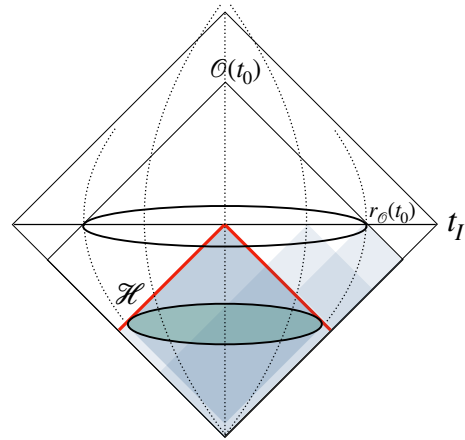}
\par\end{centering}
\protect\caption{Formation of   perturbations in holographic inflation. The geometry of the shaded region inside the horizon $\cal H$   has  a coherent quantum relationship with the external region, and curvature perturbations in relation to ${\cal O}$ are matched to external classical values at the comoving location of its boundary.  Each causal diamond surface centered on ${\cal O}$ is a 2-sphere of comoving radius  $r_{\cal O}(t_0)$ with coherent  perturbations relative to ${\cal O}$, indicated here by a filled oval. The relative potential of a comoving location  is fixed when it exits ${\cal O}$'s inflationary horizon, when  according to ${\cal O}$, it decoheres, or  ``collapses'' into a classical curvature perturbation.  Other observers have different inflationary horizons (some of their causal diamonds are also shown as shaded regions) so the system as a whole, and in particular transverse positions on $\cal H$,  remain in a superposition until $t_I$. 
\label{nested}}
\end{figure}

\begin{figure}
\begin{centering}
\includegraphics[width=\linewidth]{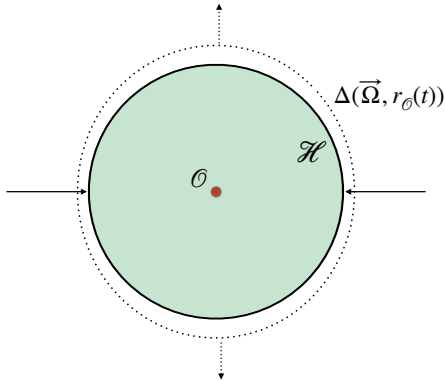}
\par\end{centering}
\protect\caption{Two dimensions of a spatial hypersurface at a single cosmic time during inflation. The shaded region lies  in the interior of the event horizon ${\cal H}$.   Incoming arrows indicate directions of incoming null information on ${\cal H}$ coming from different directions in that frame. Dotted lines indicate the coherently perturbed outgoing comoving sphere  that carries the reduced quantum state, which emerges as a  conserved classical curvature perturbation  ${\Delta}(\vec\Omega, r_{\cal O}(t))$.  The  distribution of   ${\Delta}$ on this  spherical comoving surface is correlated between all directions $\vec\Omega$, and
  matches the constraints of holographic  causal symmetries on ${\cal H}$.
\label{hypersurface}}
\end{figure}


 \subsection{Classical inflationary space-time}

An unperturbed inflationary 
universe has a 
Friedmann-Lem\^aitre-Robertson-Walker metric, with space-time interval 
\begin{equation}\label{FLRW}
ds^2 = a^2(t) [c^2d\eta^2- d\Sigma^2],
\end{equation}
where  $t$ denotes proper cosmic time for any comoving observer, $d\eta\equiv dt/ a(t)$ denotes a conformal time interval, and  $a(t)$ denotes the cosmic scale factor, determined by the equations of motion.
The  spatial 3-metric in comoving coordinates is
\begin{equation}\label{flatspace}
d\Sigma^2 = dr^2 + r^2 d\Omega^2,
\end{equation}
where the angular interval in standard polar notation is $d\Omega^2 = d\theta^2 + \sin^2 \theta d\phi^2$.
Future and past light cones from an event are defined by a null path,
\begin{equation}\label{null}
d\Sigma = \pm cd\eta.
\end{equation}
Causal diagrams for an inflationary metric are  shown in Figures  (\ref{penrose}) and (\ref{nested}).  
The end of inflation $t_I$ is taken to be the time   
 when the  expansion changes from accelerating, $\ddot a>0$, to decelerating, $\ddot a<0$.
A causal diamond  for an observer ${\cal O}$   with boundary at $t_I$ corresponds   to an interval with  equal conformal time before and after $t_I$.


The inflationary horizon $ {\cal H}$ is  an inbound null surface that
arrives at an observer ${\cal O}$ at the end of inflation. 
For our purpose, the exact choice of null surface (and $t_I$) does not matter; the important thing  is that $ {\cal H}$
forms the future boundary of a series of causal diamonds 
of nearly constant area $ 4\pi (c/H)^2$ during the slow-roll phase.

\subsection{Holographic inflation}
To calculate quantum perturbations in the classical picture,  the standard quantum model of inflation uses a straightforward extrapolation of field theory, including linearized quantum general relativity.
The holographic hypothesis holds  that this model is radically incomplete, because  the linearized approximation to quantum gravity neglects the effects of nonlocal correlations in coherent geometrical states, including information that controls  the spatial structure of quantum collapse and state reduction in transverse directions for each mode.
Estimates of the physical effects of  coherent  quantum gravity in other contexts, such as flat space-time  and black holes,  are summarized below in the Appendix. 

With  new holographic correlations  of emergent quantum gravity,
linearized general relativity no longer produces the most important scalar curvature fluctuations on the horizon scale,
so it omits the dominant effect.
The  physical difference from the  standard scenario is that  quantum geometry collapses on coherent light cone states instead of linear plane wave states (Figs. \ref{penrose}, \ref{nested} and \ref{hypersurface}). The 
relic classical curvature  perturbations, relative to an observer's world line, originate on  the  horizon. During inflation, 
they are indeterminate inside the horizon, and some of their properties are frozen outside. 
Holographic correlations within the horizon create the new features of the relic pattern discussed here.

In a semiclassical model of holographic inflation\cite{PhysRevD.99.063531},   perturbations are due to  quantum variations of holographic geometry in a nearly-uniform  background that approximates standard slow-roll inflation.
The amplitude is dominated by the new geometrical uncertainty, so the  model can omit the effect of inflaton field perturbations.
As usual, the  initial
power spectrum is nearly scale-invariant, with a small tilt that fits cosmological data with a simple effective potential.
However, new symmetries are generated on the horizon,  associated with new holographic constraints. 

In particular, collapse on the spherical horizon surface imprints large scale  correlations  on the initial state  of classical perturbations in the directional domain.
The phase correlations occur between directions around any observer, 
from the projection  of the causal diamond states of the quantum system   onto  emergent classical wave vectors (Fig. \ref{hypersurface}).
Consistency requires directionally coherent holographic correlations throughout causal diamonds.
Phase correlations have a distinctive 3D structure different from any nongaussian field:
coherent collapse occurs onto spherical horizons around every world line, rather than infinite plane waves\cite{PhysRevD.99.063531}.
The directional entanglements are most conspicuous on horizon scales similar to mode wavelength and at large angles, where the spherical curved surface of a horizon entangles different directions. 

 Features of  holographic cosmology on subhorizon scales
depend on the details of the fundamental degrees of freedom.
Ultimately, a radical  conceptual reworking of quantum geometry from holographic effects  should  lead to a  new mechanism of inflation, and may lead to new quantum effects that
extrapolates to small scales, including estimated nonlinear effects from virtual black holes\cite{Scardigli:2010gm,Banks:2018ypk}.
In a  comprehensive theory of  quantum gravity
 the new geometrical states are connected with emergence of locality,  inertial frames, angular momentum, and internal spin. 
 The simpler semiclassical model adopted here and in ref. \cite{PhysRevD.99.063531}, based on  linear perturbations of a classical background, is adequate to model the amplitude and some symmetries of new, nonlocal horizon-scale perturbations.

At  the deeper level, a full theory of emergent gravity should  account for the emergence of the cosmological  background. Measurable geometrical relationships on emergent causal diamonds are invariant, but the distinction between perturbation and
background is  observer-dependent. 
For each even frozen perturbation mode, the part that projects onto the  monopole harmonic on the sphere is absorbed into the background curvature for each observer; 
for each odd mode, the part that projects onto the dipole harmonic collapses the observer's world line to the local cosmic rest frame.
An emergent background  allows every observer  to observe  the same directional symmetries. The symmetries are 
 the same over time for each observer, aside from changes in the fractionally small time variations  from  the slowly  retreating horizon.

\subsection{Semiclassical model of  perturbations}

The premise of holographic inflation is that  nonlocal space-time indeterminacy creates coherent scalar curvature perturbations  on  horizons of the emergent inflationary metric.
 We adopt a  semiclassical model where exotic geometrical correlations on the light cone ${\cal H}$ imprint correlations on the emergent curvature of comoving world lines as they pass through ${\cal H}$, as shown in 
figures    (\ref{nested}) and (\ref{hypersurface}).

Adopting standard conventions for  linear perturbations\cite{Kadota:2005hv,Baumann:2009ds,Kamionkowski:2015yta}, denote the invariant curvature perturbation\cite{PhysRevD.22.1882} in comoving 3-space by ${\Delta}(\vec r)$,  and its spectral transform  in comoving wavenumber space $\vec k$ on surfaces of constant cosmological time by
\begin{equation}\label{complex}
\tilde{{\Delta}}(\vec k) =\int d\vec r {\Delta}(\vec r) e^{i\vec k \cdot \vec r} =  |\tilde{{\Delta}} (\vec k) |  e^{i\theta(\vec k)}.
\end{equation}
The following analysis concentrates  on  directional correlations, so  perturbations  will be described in the polar coordinates adopted for the metric, ${\Delta}(r,\theta,\phi)$.

Usually, quantum coherence is assigned and matched to the gravitational effect of plane-wave modes of amplitude $\tilde{{\Delta}} (\vec k)$ on  infinite spacelike hypersurfaces.   In the holographic model, the relic curvature perturbations are matched to  fluctuations of the quantum system projected on spherical causal diamond boundaries of the inflationary horizon ${\cal H}$. 
The fluctuations   freeze in as  differences of potential $ {\Delta}$ from the  observer when  a  comoving world line passes through the horizon. 


As discussed in the Appendix, the  physical effects of coherent states\cite{PhysRevD.99.063531}, not accounted for in  linearized gravity,  can be estimated using  
 standard classical relativity and quantum   mechanics.  
The   fractional fluctuation power  of dimensionless  perturbations on a  coherent horizon of radius $c/H$ is  given by  
\begin{equation}\label{potentialvariance}
\langle {\Delta}^2 \rangle  =  Ht_p,
\end{equation}  
where observed perturbations\cite{2013ApJS..208...20B,2013ApJS..208...19H,WMAPanomalies,Ade:2015xua,Ade:2015lrj,Array:2015xqh,Ade:2015hxq,Akrami:2018vks,Aghanim:2018eyx,Akrami:2018odb,Aylor:2017haa} have ${\Delta}^2 \sim 10^{-9}$.
Perturbations that scale like Eq. (\ref{potentialvariance}) with slowly varying $H$  produce a  nearly-scale-invariant power spectrum indistinguishable from standard cosmology\cite{Banks:2018ypk,PhysRevD.99.063531}, so they  agree with standard spectral measurements of CMB anisotropy and cosmic structure for a suitable choice of inflationary potential.  However, unlike the standard  model,  the coherent horizon has correlations among different spatial directions, and over a broad band of $k$.


\begin{figure}
\begin{centering}
\includegraphics[width=\linewidth]{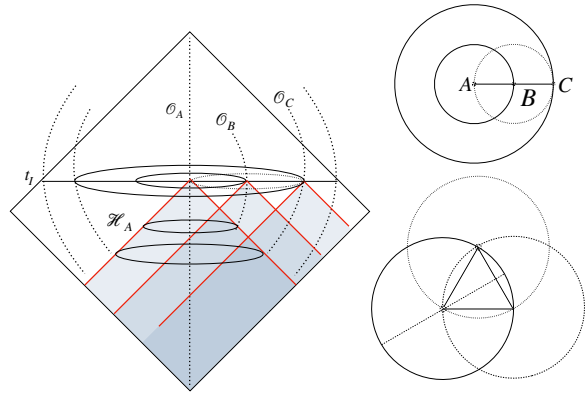}
\par\end{centering}
\protect\caption{ Spherical footprints of horizons for three world lines $A,B,C$, at  particular comoving times in the frame of $A$.  
Slices of spheres are shown at the end of inflation at the right.
Along any spatial line after this time, the sum of antipodal differences and sums must agree  for any set of tangent spheres with centers on the line. Intersections of spheres centered on each other's boundaries requires symmetries in correlations of polar values with azimuthal averages. The special case of equal size spheres is associated with polar angles $\Theta= \pm 30^\circ$, as shown at lower right.  \label{spheres}}
\end{figure}

\subsection{Correlations on spherical horizon footprints}

Even without a theory of quantum gravity,  the semiclassical model allows application of  principles governing coherent holographic emergence of a classical metric during inflation. 
The process should preserve statistical homogeneity, isotropy, and near-scale-invariance, so 
invariant scalar perturbations ${\Delta}(\vec\Omega, r_{\cal O}(t))$ freeze on nearly-spherical horizons defined by the homogeneous solution.
In any observer's frame, the   horizons around any world line  define spherical   ``footprints''  at the end of inflation (Fig. \ref{spheres}), that intersect on circles.
As usual in inflation,  smaller spheres freeze later, with coherent displacements.
Thus, smaller circles and angular positions also freeze later. 
Frozen values are coherent on circles for each sphere, so some statistical properties of $\Delta$ on equatorial circles freeze before those of smaller  circles.
The  outcome at the end of inflation is classical and determinate, so the perturbations on the spheres are not independent: in particular, the  linear sum of  polar differences or sums of curvature along any line matches the total difference or sum.

Define the   angular correlation function for   angular separation $\Theta$ for any sphere,
 \begin{equation}\label{othercirclemean}
C_{\Delta}(\Theta)\equiv \langle {\Delta}(\vec\Omega)   \langle {\Delta} \rangle_{\Theta,\vec\Omega}  \rangle_{\vec\Omega},
\end{equation}
where $ \langle \rangle_{\Theta,\vec\Omega}$  denotes an azimuthal mean on a circle at a  polar angle $\Theta$ about direction $\vec\Omega$.
This quantity is measurable on any sphere after the end of inflation. 
The holographic spherical coherence places new constraints on $C_{\Delta}(\Theta)$;
for example, we expect that frozen circles are correlated with frozen polar values only for intersections with smaller spheres that freeze out later. The
intersections of equal-radius spheres centered on each other define particular angles where consistency requires  exact, universal symmetries in the angular domain.

\subsection{Candidate symmetries}
 
We seek to  test the hypothesis  that the pattern of relic curvature   ${\Delta}(r,\theta,\phi)$ is a coherent  projection of a quantum state  on the comoving sphere $r$ when it leaves $\cal H$.  The observed classical value depends on the quantum phase of the state when it decoheres.  The semiclassical model  relates information flow in the classical world outside the horizon among  events where it affects the phase of  ${\Delta}$  when and where it decoheres,  as a function of comoving position $r,\theta,\phi$.  It allows us to 
 identify  candidate   symmetries  of  correlations in curvature perturbations in $\theta,\phi$ at a fixed $r$,  derived from causal principles proposed to govern how emergence works.
The candidate directional symmetries are based on the idea that emergent
 scalar curvature perturbations in different directions have nonlocal  relationships determined by coherent states of a quantum system with some fundamental symmetries.

\begin{figure}
\begin{centering}
\includegraphics[width=\linewidth]{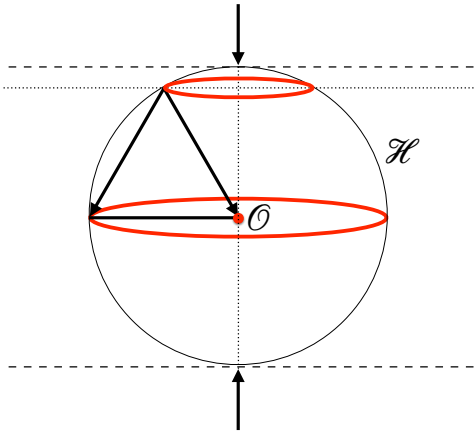}
\par\end{centering}
\protect\caption{ How azimuthal averages of  emergent perturbations are constrained by causal symmetries of state reduction.
The  2D  spherical surface of $\cal O$'s horizon is shown  in comoving 3-space at one time during inflation.
Dashed lines represent the normal surfaces to a classical emergent polar direction around the observer, the
tangent planes of incoming null surfaces on the horizon.
Since this information from polar directions along any axis reaches the observer only at the end of inflation,
it  cannot be correlated with the mean perturbation on the equator, so the
angular correlation function of curvature vanishes at $\Theta =90^\circ$.
The equilateral triangle and dotted lines show a similar causal symmetry that affects global azimuthal averages of a collapsed causal diamond at $\Theta =30^\circ$:  the  polar information    arrives at both the observer and the horizon equator at the end of inflation together with that from $\Theta =30^\circ$. \label{specialangles}}
\end{figure}


\subsubsection{Azimuthal  symmetries}

One consequence of emergent causality is that incoming  phase information  that determines  polar values of potential along any given axis on the horizon only reaches the equatorial plane at the end of inflation, so it cannot  affect the potential for points in that plane  (see Fig. \ref{specialangles}).  
As shown in the examples given in the Appendix, the essential nature of coherent  holographic uncertainty is that it is entirely transverse to the direction of propagation. 

 Uncorrelated incoming polar information   produces the following {\it exact global equatorial symmetry} of relic curvature on each sphere around an observer:  
\begin{equation}\label{globalequatorial}
\langle {\Delta}(\vec\Omega)   \langle {\Delta} \rangle_{\perp\vec\Omega}  \rangle_{\vec\Omega}= 0.
\end{equation}
where $\langle  \rangle_{\perp\vec\Omega}$ denotes the azimuthal mean on the great circle normal to direction $\vec\Omega$, and
$\langle  \rangle_{\vec\Omega}$ denotes an average over all directions $\vec\Omega$.
Since by definition the point average vanishes,
$\langle {\Delta}(\vec\Omega)\rangle_{\vec\Omega}  =0$,
Eq. (\ref{globalequatorial})  follows  if  {\it the average $\langle {\Delta} \rangle_{\perp\vec\Omega}$ on each  great circle is uncorrelated with the sum of its polar values,  ${\Delta}(\vec\Omega)+{\Delta}(-\vec\Omega)$}.

Such an exact   equatorial symmetry would follow if there were exactly odd point parity, or if
 all great circles have vanishing mean ($ \langle {\Delta} \rangle_{\perp\vec\Omega}=0$). 
However, it does not require either of these to  hold: it specifies an orthogonal  rather than an antipodal relationship, and can hold exactly even if  antisymmetry is not exact.  
 It  arises if  {\it perturbations on orthogonal axes  are always independently generated}.
Eq. (\ref{globalequatorial}) is equivalent to
\begin{equation}\label{globalequatorial2}
C_{\Delta}(90^\circ)= 0.
\end{equation}

This  symmetry does not apply in the standard scenario.   In  that case,   equatorial azimuthal symmetry occurs only by chance, and only approximately: in a harmonic description, it requires  a global ``conspiracy'' of phases  of even-parity modes.  From this point of view, an  exact equatorial azimuthal symmetry is the most simple and direct signature of emergent causal structure in holographic inflation.

As  shown in Fig. (\ref{specialangles}),  certain assumptions about how emergence works  could also  lead to  an  equilateral azimuthal symmetry.
Consider the correlations of a polar value with points on its $30^\circ$ azimuthal circle, both again determined by incoming phase information from the polar direction.  Data from any polar point and from its $30^\circ$ circle on the same sphere  reach both the observer and the equator at the end of inflation. If the azimuthal mean is independent of its polar value,
the global mean  vanishes, so 
\begin{equation}\label{corr30}
C_{\Delta}(30^\circ)=  0.
\end{equation}  
   
In a multipole expansion,  a $C_{\Delta}( 30^\circ)=0$ symmetry constrains  odd as well as even spherical harmonics.  In particular, 
 a nonzero dipole contributes $C_{\Delta}( 30^\circ)\ne 0$: thus, for $C_{\Delta}( 30^\circ)=0$ to appear on a dipole-subtracted sky,  it   must apply in  the  frame of an observer where  the dipole vanishes.  For this to occur independent of  the cosmic rest frame,  the intrinsic dipole, as viewed in the local matter rest frame, should itself vanish due a new symmetry that governs how  the cosmic comoving frame emerges from a quantum system.   The complementary azimuthal mean  at $\Theta =180^\circ-30^\circ=150^\circ$  could be small but nonzero, depending on  the origin of the local cosmic rest frame and global parity violation.

 \subsubsection{Constant variance on great  circles}

The global equatorial symmetry still allows  mean values of ${\Delta}$ on great circles to vary according to random incoming polar information associated with each axis.  A separate symmetry may be associated with nonlocal correlations of variance normal to each axis. It is possible that {\it the  variance on great circles is a constant}, that is, it  depends only on basic physics, rather than being a random variable.

This symmetry appears in  a toy model described in the Appendix.  In  (Eq. \ref{3D}), the commutator on the left hand side that is responsible for the exotic quantum fluctuation has projected directional components orthogonal to the  components on the right hand side, related by the antisymmetric tensor $\epsilon_{ijk}$,  and leads to an uncertainty,  Eq. (\ref{spinuncertainty}). Thus, an eigenstate with $\delta{\tau}_3=0$ has an irreducible uncertainty in the sum of orthogonal components, $ \langle\delta{\tau}_1^2 + \delta{\tau}_2^2\rangle$.

A similar symmetry applied to curvature  during inflation would relate  perturbation power in orthogonal directions.
A  possible consequence is that a directional average in the  plane normal  to any  direction, say $3$, obeys the same relation:
\begin{equation}\label{diamondcurvature}
\langle {\Delta}^2 \rangle_{\perp 3 } = \langle\delta{\tau}_1^2 + \delta{\tau}_2^2\rangle/ {\tau} ^2 \ge  \langle {\Delta}^2 \rangle.
 \end{equation}

Suppose that   cosmology   preserves the symmetry of statistical isotropy, that is, statistical quantities are independent of direction.  
In order for Eq. (\ref{diamondcurvature})  to hold for any direction in a single statistically isotropic distribution, the bound must saturate  in all directions:   if any direction were to exceed the overall average, another direction would need to have less than the lower bound.
This leads to a  symmetry of the azimuthal average variance:  
\begin{equation}\label{equatorvariance}
\langle {\Delta}_\perp^2\rangle_\phi \equiv \int d\phi' [ {\Delta}(\phi', \theta'=\pi/2)]^2 = \langle {\Delta}^2\rangle,
\end{equation}
for any orientation of polar coordinates $(\theta',\phi')$.
That is, {\it the  variance of curvature perturbations  on any great circle is equal to  the  variance for the whole sky.}

The  symmetry represented by Eq. (\ref{equatorvariance})  
does not hold for  random-phase noise in standard quantum inflation.
It arises from the rotational symmetry of  the  quantum system, and the holographic absence of one  independent rotational degree of freedom.
Thus,  it can be used to differentiate the distribution from the random gaussian noise predicted in the  standard model.

As usual,  there is  a zero mean variation averaged  over all directions $\vec\Omega$,
\begin{equation}\label{zeroplane}
\langle {\Delta}\rangle_{\vec\Omega} \equiv \int \int d\phi d\theta \sin(\theta) {\Delta}(\phi,\theta) =  0,
\end{equation}
so  ${\Delta}$ can be decomposed into  a linear sum of spherical harmonic components $Y_\ell^m$.
To satisfy the symmetry of great circle variance  (Eq. \ref{equatorvariance}) they must have a coherent  relationship that appears as a conspiracy of alignments and amplitudes.
It is useful to illustrate with an  example that uses a few low order spherical harmonics:  octopole ($\ell=3$), quadrupole ($\ell=2$), and dipole ($\ell=1$).  

Suppose there is a  dipole  aligned along the  $z$ axis, with
${\Delta}_1\propto \cos \theta$. 
It represents the intrinsic dipole  of curvature  in the polar direction, as viewed in the local cosmic rest frame.
By itself, it  obeys Eq. (\ref {equatorvariance}) for  all great circles that pass through the pole, with normal directions in the $x,y$ plane.

 For the dipole alone, the variance on the equator $\theta= \pi/2$ vanishes so it does not satisfy Eq. (\ref {equatorvariance}).
The other multipole moments must organize around this direction to have  perturbations satisfying Eq. (\ref {equatorvariance})   in all directions.
The fit improves if we add   to  the dipole  a  precisely aligned sectoral octopole ($\ell=|m|=3$),  with angular  dependence
 ${\Delta}_3\propto e^{3i \phi}\sin^3 \theta$, and a precisely aligned sectoral quadrupole ($\ell=|m|=2$), with angular  dependence
  ${\Delta}_2\propto e^{2i \phi}\sin^2 \theta$.
 The dipole variance vanishes along the equator, while the  octopole and quadrupole variances    are maximized for an equatorial great circle. 
 To create a pattern consistent with Eq. (\ref {equatorvariance}) in all directions requires correlations and alignments of higher multipoles, although the symmetry could apply over only a limited range of  $\ell$, depending on the angular coherence scale of correlations. 
 Some manifestations of  multipole alignment  in the CMB are  outlined below.

\subsubsection{Antipodal anticorrelation  and parity violation}

The inflationary horizon could also display  antipodal anticorrelation, a tendency of opposite points in the sky to have opposite signs.   
It occurs in quantum models  of eternal black hole horizons\cite{Hooft:2016cpw,Hooft:2016itl,Hooft2018}, where antipodes on the horizon are actually identified,  and time-reversed conjugate particle states are entangled at opposite spatial poles, even on macroscopic scales.
 In flat space-time,  a toy model of quantum nonlocality (see the Appendix below) derives point-parity antisymmetry   directly from the Dirac light cone structure (Eq. \ref{antiDelta}), which carries over into geometrical  operators:
\begin{equation}\label{antitime}
\hat {\tau} (x_\kappa) = - \hat {\tau} (-x_\kappa).
\end{equation}
 In our model of emergent inflation, the antipodal outgoing  states of geometry on the horizon become antipodal curvature perturbations of the emergent metric. As discussed above,  antipodal relationships are entangled with the emergence of local cosmic rest frame.

In standard inflation,  any measured asymmetry in antipodal perturbation power  is entirely due to ``cosmic variance'' from the zero expected value of the ensemble.
In the range of possible realizations, the bulk of the  point-parity antisymmetry (or symmetry) is almost always contributed by a small number of harmonic modes on  large angular scales, especially the intrinsic dipole and quadrupole.

In the holographic model, as in black holes,  a coherent holographic horizon can provide a global constraint on even and odd  parity relationships between opposite points in the sky for any realization.  Indeed  there may be no cosmic variance in the ratio of odd to even perturbation power; its value may be set by  underlying asymmetric physics.
 In general, violation is not confined to low-$\ell$ modes.

To start with, consider the extreme case:    frozen metric perturbations with exact antisymmetry similar to eternal black holes, so that
the  curvature in direction $\vec\Omega$ satisfies 
\begin{equation}\label{anticurvature}
{\Delta}(\vec\Omega)= - {\Delta}(-\vec\Omega),
\end{equation}
an exactly odd point parity of  perturbations.
Microscopic spookiness  is manifested macroscopically: points in   opposite directions  ``know about each other'' like nearby ones do.
Unlike the standard picture, 
{\it  spooky  correlation does not separate scales: it  applies to fine-grain angular detail (that is, high resolution $\ell>>1$),  even at large angular separation}. 
In this extreme example, the all-sky distribution has an antipodal variance equal and opposite to the  single-point variance:
\begin{equation}\label{meananticurvature}
\langle {\Delta}(\vec\Omega) {\Delta}(-\vec\Omega) \rangle= - \langle {\Delta}^2 \rangle.
\end{equation}

We should  allow for the possibility of a less extreme imbalance of odd and even perturbations.
 In cosmology, as in a realistic time-asymmetric black hole,  an exact antipodal antisymmetry can be broken by a background system that is not in a time-symmetric equilibrium or ground state.
The exact antipodal antisymmetry of flat-space-time may not  apply  for  inflationary perturbations, since 
 the classical inflationary  background  breaks the time-displacement and boost symmetries of classical relativity, and the vacuum matter fields that couple to gravity  also in general violate parity.
The magnitude of time-direction symmetry  breaking is related to how much the  inflationary solution departs from that of  the maximally symmetric de Sitter inflationary solution.
One  direct measure of  time asymmetry  comes from the slight tilt of the power spectrum  from exact scale invariance, measured\cite{Aghanim:2018eyx,Akrami:2018vks}  to be $1-n_S = 0.035\pm0.004$, which arises from  the small fractional decrease in the expansion rate during each $e$-folding of inflation\cite{PhysRevD.99.063531}.
We  adopt a cosmological  symmetry breaking parameter ${\cal E}<1$, so that  Eq. (\ref{meananticurvature})   becomes
\begin{equation}\label{antiprediction}
\langle {\Delta}(\vec\Omega) {\Delta}(-\vec\Omega) \rangle = - \langle {\Delta}^2 \rangle  (1-{\cal E}),
\end{equation}
where the ratio of even to odd perturbation power is ${\cal E}$. 

In general, ${\cal E}$ depends on  angular wave number $\ell$. The fine grain character of parity asymmetry is best revealed by  harmonic analysis. An analysis of temperature anisotropy can be found for example in Fig. 25  of ref. \cite{Akrami:2019bkn}, where the value of ${\cal E}$ for temperature anisotropy at each $\ell$ is denote by $R_{TT}$.  It  shows a detection of ${\cal E}_{T}\approx 0.3$ at $\ell$ of order a few, and a significant detection of power asymmetry  up to $\ell \approx 30$.

 .

\section{Patterns in   CMB anisotropy}
It is useful to define  measures of symmetry that can be used for empirical  tests  based on  
CMB anisotropy. The symmetries of curvature  lead to simple predictions that can be used as the basis for tests  on large angular scales, $\Theta= {\cal O}(1)$.
The most distinctive signature of exotic symmetries is that they constrain this large-angle pattern even at high resolution, $\ell>>1$.

On large angular scales, CMB  temperature anisotropy is dominated by perturbations within a relatively thin sphere at the epoch of last scattering, with some secondary effects from the intervening volume\cite{1967ApJ...147...73S,Hu:2001bc,Wright:2003ig}.
The direct effect of scalar perturbations  on anisotropy is described by the Sachs-Wolfe approximation\cite{1967ApJ...147...73S}; in this approximation, the pattern of $\delta T$ preserves the pattern (and symmetries) of $\Delta$.  However, the fine-grained temperature distribution for harmonics $\ell>>1$  is significantly  affected by Doppler motions, and these must be included in tests of precise symmetries, 
even at large angular separation, $\Theta\sim {\cal O}(1)$. 
Because the pattern of Doppler anisotropy for each 3D wave vector differs from the scalar pattern, a
linear algorithm may be used  to reconstruct the pattern of the primordial potential from measurements of temperature and/or polarization
\cite{2005PhRvD..71l3004Y,Dorn:2014rua}, although this has not yet been done for an all-sky CMB map.



\subsection{Symmetries of the  correlation function}
The  two-point correlation function defined above (Eq. \ref{othercirclemean}) can also be written in terms of an empirical estimator
\begin{equation}\label{corrdef}
C_\Delta(\Theta) = \langle{\Delta}_a {\Delta}_b \rangle_{\angle ab = \Theta},
\end{equation}
an all-sky average  over all pairs of points $a,b$  at angular separation $\angle ab = \Theta$.
Relics of primordial symmetries on the inflationary horizon survive in $\Delta(\theta,\phi)$ at last scattering, and can be reconstructed from measurements.

In the particular case of angular separation $\Theta= 90^\circ$, the Doppler contribution  vanishes.  The reason is that for any 3D mode, the velocity perturbation vanishes for directions in the plane.  This also applies to the contribution from our own velocity, the kinematic dipole.
Thus  the particular case of equatorial azimuthal symmetry (Eq. \ref{globalequatorial2}) is particularly robust:  a reconstruction is not necessary, and the symmetry survives  in the temperature. 
An exact global equatorial symmetry of curvature (Eq. \ref{globalequatorial2})
{\it  implies  exactly vanishing temperature correlation at 90 degrees}: 
\begin{equation}\label{corrperp}
C_{ T}(90^\circ)= C_{\Delta}(90^\circ)=   0.
\end{equation}
This exact property of the correlation function applies even at high angular resolution in a holographic model with the exact symmetry given by Eq. (\ref{globalequatorial}); it can occur by chance in some realizations of a conventional cosmology, but only very rarely to a very high precision.

Causal constraints on correlation of polar and azimuthal information  (Fig. \ref{specialangles})  could also lead to exact symmetries at other angles, for example at $30^\circ$
(Eq. \ref{corr30}), but in  this case,  a reconstruction is necessary; the exact symmetry does not survive in $\delta T$.

The distribution on the sky is often described in terms of spherical harmonics $A_{\ell m}$.
The angular power spectrum
\begin{equation}
C_\ell= \frac{1}{2\ell+1}
\sum_{m= -\ell}^{m=+\ell} | A_{\ell m}|^2
\end{equation} 
has the same information as the angular correlation function;
contributions of odd and even spherical harmonics to the correlation function are given by the  standard formula  (e.g., \cite{WMAPanomalies,Schwarz:2015cma,Hinshaw_1996})
 \begin{equation}\label{harmonicsum}
C(\Theta) = \frac{1}{4\pi}\sum_\ell (2\ell +1) {\cal C}_\ell P_\ell (\cos \Theta),
\end{equation}
where $P_\ell $  are the Legendre polynomials, with the property that $P_\ell(0) = 0  $  for odd  $\ell$, $P_\ell(0) \ne  0 $ for even $\ell$.

In the harmonic description,  a simple exact symmetry in the angular domain,  such as Eq. (\ref{corrperp}), appears as a conspiracy of many spherical harmonic coefficients. The conspiracy extends to high $\ell$, even for symmetries at large $\Theta$. In particular, agreement with 
$C(90^\circ)=   0$ requires a sum of even mode contributions to vanish at high precision up to the  map resolution $\ell$, which is very unlikely if they are drawn independently from random distributions as in the standard picture.

\subsection{Interpretation of  well known  CMB anomalies}
The  holographic model  provides  a unified physical interpretation that accounts in a general way for several long-studied features of measured CMB anisotropy at low $\ell$ that are statistically anomalous in the standard model\cite{WMAPanomalies,Ade:2015hxq,Akrami:2019bkn,2017MNRAS.472.2410A,2012MNRAS.419.3378A,2015MNRAS.449.3458C,Schwarz:2015cma}.  
 We first summarize a proposed interpretation of some well known empirical anomalies, then suggest  more sharply defined theory-motivated tests that can differentiate holographic inflation from the standard picture.

{\it
 1. Axes defined by the quadrupole and  octopole  are  closely aligned.}
 The  {\it WMAP} all-sky maps\cite{2013ApJS..208...20B,2013ApJS..208...19H,WMAPanomalies} revealed
 a remarkably close agreement in direction for quadrupole ($\ell=2$) and octopole ($\ell =3$) harmonics. 
  The aligned direction is  defined by the
axis that maximizes the sum of the squares of  $a_{\ell,\ell}$ and $a_{\ell,-\ell}$ spherical harmonic coefficients, that is,  maximizes polar asymmetry.  A variety of studies have confirmed the close alignment  to be
 highly unlikely in the standard model\cite{Schwarz:2015cma}. 
In  our model, as discussed above,  the principal axes of  harmonics have to be aligned to satisfy the 
constraints imposed by a holographic information deficit with rotational symmetry, such as  constant variance on great circles (Eq. \ref{equatorvariance}). Note that the alignments depend on the $A_{\ell m}$'s but not the $C_\ell$'s, so this is a higher-order symmetry not captured by symmetries of $C(\Theta)$. 

Since the alignment axis is associated with  physical primordial modes  in three dimensions, whose orientation is  correlated  on different scales,  the model could  account for  why  ``secondary''  Integrated Sachs-Wolfe (ISW \cite{1967ApJ...147...73S,Hu:2001bc,Wright:2003ig}) contributions from gravitational effects in the intervening volume do not spoil the  precise alignment.

{\it
2. The  two-point temperature correlation function is small at  large angular separation} \cite{Ade:2015hxq,Akrami:2019bkn,Schwarz:2015cma}.
An unexpected lack of large angle correlation power has been apparent since the first measurements with COBE\cite{Hinshaw_1996}.  

In the  {\it WMAP} analysis of  $C(\Theta)$,  based  on 7 years of data\cite{WMAPanomalies},   the authors comment on the (true) fact that there is no significant conflict with the standard random-phase scenario, and no significant deficit of large-scale power: 
``$C(\Theta)$ lies within the 95\% confidence range of the best-fit $\Lambda$CDM model for all $\Theta$, as determined by Monte Carlo simulations. This supports the conclusion that there is no statistically significant lack of large-scale power on the full sky.''
At the same time,   compared with a standard ensemble of random-phase CMB realizations, $C(\Theta)$  is both anomalously close to zero around 90 degrees, and anomalously negative near 180 degrees.  
  
  By contrast,
the published  {\it WMAP} plot of $C_T(\Theta)$ appears to agree remarkably well with the simple and exact global equatorial symmetry  (Eq. \ref{corrperp}) of holographic inflation:
the  measured $C_T(\Theta)$   appears to have  a value at 90 degrees consistent with zero. 

{\it 3. The quadrupole  and other even harmonics are smaller than expected}.
As discussed above, an excess of odd over even fluctuation power \cite{Ade:2015hxq,Akrami:2019bkn,Schwarz:2015cma},  measured    in harmonic decomposition,  shows significant anomalous antipodal anticorrelation on angular scales much smaller than the dipole, which also appears as a significant  negative correlation in $C_T(\Theta)$ near 180 degrees.
Both 
are  interpreted here as  direct  signatures of  antipodal anticorrelation on the horizon, with ${\cal E} <1$ over a wide range of $\ell$ (Eq. \ref {antiprediction}).

\subsection{New tests }

In holographic inflation, 
precision tests of symmetries are possible on large angular scales without the usual cosmic-variance penalty on significance:
 predictions for  symmetries  are not influenced by the random variables of realizations as they are in the standard scenario.
This feature allows for powerful parameter-free comparisons with standard theory. 

\subsubsection{ Global azimuthal symmetries}

  The exact value of 
$C(\Theta)$  has no  particular significance in the standard scenario, so a  very close agreement with zero can indicate a likelihood with a strong preference for the holographic theory.
Such a precise null test of equatorial symmetry   (Eq. \ref{corrperp})  will require more  attention to specific biases of foreground subtraction and masking than needed in large scale tests of the standard picture, whose predictions have a large scatter from cosmic variance.
 
Published {\it WMAP} and {\it Planck}  plots disagree in detail.  In the case of the  {\it WMAP} ILC all-sky map\cite{WMAPanomalies}, $C_T(90^\circ)$ appears to be consistent with zero.  In  the case of {\it Planck} \cite{Ade:2015hxq,Akrami:2019bkn}, four different  foreground removal techniques agree with each other, and 
$C_T(90^\circ)$  differs slightly but significantly from zero in all of them.  Recently, these maps were  re-analyzed (\cite{Hagimoto_2020}) with uniform masking. The result of this analysis was that the least biased measurements of $C(\Theta)$  from the two satellites and most of the different foreground removal techniques are in good agreement with each other near $\Theta=90^\circ$, and fall into a remarkably narrow range  around $C(90^\circ)=0$.  A new, more precise likelihood comparison  with  standard theory  should be possible with closer attention to foregrounds, and with a  reconstruction of $C_\Delta$, perhaps including polarization data\cite{2005PhRvD..71l3004Y,Dorn:2014rua}.

\subsubsection{Intrinsic Dipole}

In emergent cosmology, it is plausible that (unlike the standard scenario) there is a vanishing intrinsic dipole, because  enclosed matter and its bounding CMB horizon share a common causal past (see Fig. \ref{nested}):  they are bounded by the  same causal diamond, whose boundary lies at the end of inflation.
As the classical metric emerges from a quantum system,  the cosmic rest frame defined by the sum of cosmic matter within the horizon (that is, into which its geometrical position state collapses) should be  the same as the frame defined by the bounding horizon.

This symmetry is testable in principle. 
The intrinsic dipole can be  estimated  by measuring and subtracting the contribution from galaxy flows on very large scales ($>200{\rm Mpc}$) that account for  the dipole contributed by  peculiar velocity. If the intrinsic dipole vanishes, the prediction is that the motional dipole  agrees with the sky dipole.  Because  the first order dipole induced by motion is so much larger than even the standard expected primordial curvature-induced  intrinsic dipole,  a precision test is difficult in practice: it requires a comprehensive  survey of precise cosmic distances ({\it e.g.}, ref. \cite{Qin:2018hue}).  

In the case  that  the intrinsic dipole does not vanish,  it must still  ``conspire'' with other odd harmonics to be consistent with  great-circle-variance symmetry, as discussed above.
In this case,  constant great-circle variance allows another test:  it should be possible to uniquely reconstruct both the axis and  amplitude of the dipole from measured maps.

\subsubsection{Future Holometer experiments}

Nonlocal coherent fluctuations on causal diamonds, as posited here for inflation, should also affect  light propagating between massive bodies in flat space on macroscopic scales\cite{Hogan:2010zs,Kwon:2014yea,Hogan:2015kva,Hogan:2015b,Hogan:2016,Verlinde:2019xfb}.
It has recently become possible to probe these coherent
Planck scale  fluctuations directly in laboratory experiments, using  interferometers with signal measurement bandwidth comparable with  their free spectral range.  

If displacements on causal diamonds in the laboratory are governed by the same uncertainty  as that used for holographic cosmological perturbations (Eq. \ref{potentialvariance}),  a differential measurement of  light paths of length $L$ displays fractional  fluctuations on timescale $L/c$ of order
\begin{equation}\label{fluctuations}
\langle\delta L^2\rangle/L^2  \approx L l_P.
\end{equation}
They may be observable as cross correlations,  with Planck scale power spectral density, between signals in interferometers
\cite{Hogan:2010zs,Kwon:2014yea,Hogan:2015kva,Hogan:2015b,Hogan:2016,Verlinde:2019xfb}.
The effect on the signal depends on the directional structure of fluctuations,  and the spatial structure of the light paths.

The results of such experiments depend  on symmetries of the apparatus, which control its response to coherent fluctuations of causal diamonds.
It  has been demonstrated experimentally that there is no such effect for  some configurations of the light paths.  The most sensitive published measurement\cite{holoshear}   constrains coplanar, radial quadrupolar fluctuations  to  more than an order of magnitude less than  the  amplitude in Eq. (\ref{fluctuations}).  Current experiments are studying the possibility of purely rotational or transversal fluctuations\cite{Hogan:2016}.  Future experiments in different configurations,  including light paths extending in three dimensions,  could  respond to holographically entangled fluctuations of light cones  similar to that posited here for  inflationary horizons.  A detection of  cosmological spookiness could provide both motivation and design guidance for future experiments.

\section{Conclusion}

The  simple  geometrical symmetries  proposed here are examples of holographic correlations: they apply to the entire sky  on all scales,  and reduce the independence of perturbations in different directions.
The primordial structure of curvature perturbations also
survives today  in three dimensional large scale structure:   the  structural  pattern of the galaxy distribution caused by spooky entanglement  should display  the same exotic  angular correlations as the  CMB pattern\cite{PhysRevD.99.063531}.
The  candidate symmetries of holographic correlations analyzed here are predicted to  appear  clearly in the CMB on large scales today   largely because the last scattering surface approximates a 2-sphere, like the primordial horizon where the correlations originated. For this reason,  CMB anisotropy at present provides the most direct tests of the holographic-inflation hypothesis.

Directional symmetries of emergent perturbations  allow the formulation of new predictions accessible to test at large angular scales with existing data. The new correlations arise from entangled relationships at  large separations ($\Theta$ of order unity in the angular domain), but also in fine grained structural detail ($\ell>> 1$ in the wavenumber domain) --- a  feature distinctly absent in the standard scenario.

An important  new  feature of  holographic inflation is that its precise symmetries allow predictions for  measurements in the angular domain with no cosmic variance, so  large-angle  tests  can  achieve  unaccustomed power.
It is remarkable that  generic, holographically-motivated  symmetries approximately account  for some well-known so-called ``anomalies'' of  CMB anisotropy.  Moreover, some new predictions are unusually specific: for example, emergent causality leads to an exact  value for the angular correlation function,  $C_T(90^\circ)=0$, that has no particular significance  in the standard picture. These predictions motivate new, specifically targeted combined analyses of the {\it WMAP} and {\it Planck} data, as well as new measurements of  polarization over a large fraction of the sky, that can further reduce systematic errors from astrophysical foregrounds.

Precise confirmation  of these emergent symmetries 
would lend support to the hypothesis that 
primordial scalar curvature perturbations originate mainly from   holographically  coherent  quantum gravity, rather than a  conventional system of quantum fields with linearized gravity. 
The observed perturbation pattern in that case is then  a direct relic of the deeper quantum system, and its symmetries can be used to constrain  theories of  emergent space-time.

  \begin{acknowledgments}
  I am grateful to T. Banks,  O. Kwon and S. Meyer for  useful discussions and comments.
 This work was supported by the Department
of Energy at Fermilab under Contract No. DE-AC02-07CH11359.  
\end{acknowledgments}

\bibliography{CMBpattern}

\begin{thebibliography}{81}%
\makeatletter
\providecommand \@ifxundefined [1]{%
 \@ifx{#1\undefined}
}%
\providecommand \@ifnum [1]{%
 \ifnum #1\expandafter \@firstoftwo
 \else \expandafter \@secondoftwo
 \fi
}%
\providecommand \@ifx [1]{%
 \ifx #1\expandafter \@firstoftwo
 \else \expandafter \@secondoftwo
 \fi
}%
\providecommand \natexlab [1]{#1}%
\providecommand \enquote  [1]{``#1''}%
\providecommand \bibnamefont  [1]{#1}%
\providecommand \bibfnamefont [1]{#1}%
\providecommand \citenamefont [1]{#1}%
\providecommand \href@noop [0]{\@secondoftwo}%
\providecommand \href [0]{\begingroup \@sanitize@url \@href}%
\providecommand \@href[1]{\@@startlink{#1}\@@href}%
\providecommand \@@href[1]{\endgroup#1\@@endlink}%
\providecommand \@sanitize@url [0]{\catcode `\\12\catcode `\$12\catcode
  `\&12\catcode `\#12\catcode `\^12\catcode `\_12\catcode `\%12\relax}%
\providecommand \@@startlink[1]{}%
\providecommand \@@endlink[0]{}%
\providecommand \url  [0]{\begingroup\@sanitize@url \@url }%
\providecommand \@url [1]{\endgroup\@href {#1}{\urlprefix }}%
\providecommand \urlprefix  [0]{URL }%
\providecommand \Eprint [0]{\href }%
\providecommand \doibase [0]{http://dx.doi.org/}%
\providecommand \selectlanguage [0]{\@gobble}%
\providecommand \bibinfo  [0]{\@secondoftwo}%
\providecommand \bibfield  [0]{\@secondoftwo}%
\providecommand \translation [1]{[#1]}%
\providecommand \BibitemOpen [0]{}%
\providecommand \bibitemStop [0]{}%
\providecommand \bibitemNoStop [0]{.\EOS\space}%
\providecommand \EOS [0]{\spacefactor3000\relax}%
\providecommand \BibitemShut  [1]{\csname bibitem#1\endcsname}%
\let\auto@bib@innerbib\@empty
\bibitem [{\citenamefont {Weinberg}(2008)}]{Weinberg:2008zzc}%
  \BibitemOpen
  \bibfield  {author} {\bibinfo {author} {\bibfnamefont {Steven}\ \bibnamefont
  {Weinberg}},\ }\href {http://www.oup.com/uk/catalogue/?ci=9780198526827}
  {\emph {\bibinfo {title} {{Cosmology}}}}\ (\bibinfo  {publisher} {Oxford
  University Press},\ \bibinfo {year} {2008})\BibitemShut {NoStop}%
\bibitem [{\citenamefont {{Bennett}}\ \emph {et~al.}(2013)\citenamefont
  {{Bennett}}, \citenamefont {{Larson}}, \citenamefont {{Weiland}},
  \citenamefont {{Jarosik}}, \citenamefont {{Hinshaw}}, \citenamefont
  {{Odegard}}, \citenamefont {{Smith}}, \citenamefont {{Hill}}, \citenamefont
  {{Gold}}, \citenamefont {{Halpern}}, \citenamefont {{Komatsu}}, \citenamefont
  {{Nolta}}, \citenamefont {{Page}}, \citenamefont {{Spergel}}, \citenamefont
  {{Wollack}}, \citenamefont {{Dunkley}}, \citenamefont {{Kogut}},
  \citenamefont {{Limon}}, \citenamefont {{Meyer}}, \citenamefont {{Tucker}},\
  and\ \citenamefont {{Wright}}}]{2013ApJS..208...20B}%
  \BibitemOpen
  \bibfield  {author} {\bibinfo {author} {\bibfnamefont {C.~L.}\ \bibnamefont
  {{Bennett}}}, \bibinfo {author} {\bibfnamefont {D.}~\bibnamefont {{Larson}}},
  \bibinfo {author} {\bibfnamefont {J.~L.}\ \bibnamefont {{Weiland}}}, \bibinfo
  {author} {\bibfnamefont {N.}~\bibnamefont {{Jarosik}}}, \bibinfo {author}
  {\bibfnamefont {G.}~\bibnamefont {{Hinshaw}}}, \bibinfo {author}
  {\bibfnamefont {N.}~\bibnamefont {{Odegard}}}, \bibinfo {author}
  {\bibfnamefont {K.~M.}\ \bibnamefont {{Smith}}}, \bibinfo {author}
  {\bibfnamefont {R.~S.}\ \bibnamefont {{Hill}}}, \bibinfo {author}
  {\bibfnamefont {B.}~\bibnamefont {{Gold}}}, \bibinfo {author} {\bibfnamefont
  {M.}~\bibnamefont {{Halpern}}}, \bibinfo {author} {\bibfnamefont
  {E.}~\bibnamefont {{Komatsu}}}, \bibinfo {author} {\bibfnamefont {M.~R.}\
  \bibnamefont {{Nolta}}}, \bibinfo {author} {\bibfnamefont {L.}~\bibnamefont
  {{Page}}}, \bibinfo {author} {\bibfnamefont {D.~N.}\ \bibnamefont
  {{Spergel}}}, \bibinfo {author} {\bibfnamefont {E.}~\bibnamefont
  {{Wollack}}}, \bibinfo {author} {\bibfnamefont {J.}~\bibnamefont
  {{Dunkley}}}, \bibinfo {author} {\bibfnamefont {A.}~\bibnamefont {{Kogut}}},
  \bibinfo {author} {\bibfnamefont {M.}~\bibnamefont {{Limon}}}, \bibinfo
  {author} {\bibfnamefont {S.~S.}\ \bibnamefont {{Meyer}}}, \bibinfo {author}
  {\bibfnamefont {G.~S.}\ \bibnamefont {{Tucker}}}, \ and\ \bibinfo {author}
  {\bibfnamefont {E.~L.}\ \bibnamefont {{Wright}}},\ }\bibfield  {title}
  {\enquote {\bibinfo {title} {{Nine-year Wilkinson Microwave Anisotropy Probe
  (WMAP) Observations: Final Maps and Results}},}\ }\href {\doibase
  10.1088/0067-0049/208/2/20} {\bibfield  {journal} {\bibinfo  {journal} {The
  Astrophysical Journal Supplement Series}\ }\textbf {\bibinfo {volume}
  {208}},\ \bibinfo {eid} {20} (\bibinfo {year} {2013})},\ \Eprint
  {http://arxiv.org/abs/1212.5225} {arXiv:1212.5225 [astro-ph.CO]} \BibitemShut
  {NoStop}%
\bibitem [{\citenamefont {{Hinshaw}}\ \emph {et~al.}(2013)\citenamefont
  {{Hinshaw}}, \citenamefont {{Larson}}, \citenamefont {{Komatsu}},
  \citenamefont {{Spergel}}, \citenamefont {{Bennett}}, \citenamefont
  {{Dunkley}}, \citenamefont {{Nolta}}, \citenamefont {{Halpern}},
  \citenamefont {{Hill}}, \citenamefont {{Odegard}}, \citenamefont {{Page}},
  \citenamefont {{Smith}}, \citenamefont {{Weiland}}, \citenamefont {{Gold}},
  \citenamefont {{Jarosik}}, \citenamefont {{Kogut}}, \citenamefont {{Limon}},
  \citenamefont {{Meyer}}, \citenamefont {{Tucker}}, \citenamefont
  {{Wollack}},\ and\ \citenamefont {{Wright}}}]{2013ApJS..208...19H}%
  \BibitemOpen
  \bibfield  {author} {\bibinfo {author} {\bibfnamefont {G.}~\bibnamefont
  {{Hinshaw}}}, \bibinfo {author} {\bibfnamefont {D.}~\bibnamefont {{Larson}}},
  \bibinfo {author} {\bibfnamefont {E.}~\bibnamefont {{Komatsu}}}, \bibinfo
  {author} {\bibfnamefont {D.~N.}\ \bibnamefont {{Spergel}}}, \bibinfo {author}
  {\bibfnamefont {C.~L.}\ \bibnamefont {{Bennett}}}, \bibinfo {author}
  {\bibfnamefont {J.}~\bibnamefont {{Dunkley}}}, \bibinfo {author}
  {\bibfnamefont {M.~R.}\ \bibnamefont {{Nolta}}}, \bibinfo {author}
  {\bibfnamefont {M.}~\bibnamefont {{Halpern}}}, \bibinfo {author}
  {\bibfnamefont {R.~S.}\ \bibnamefont {{Hill}}}, \bibinfo {author}
  {\bibfnamefont {N.}~\bibnamefont {{Odegard}}}, \bibinfo {author}
  {\bibfnamefont {L.}~\bibnamefont {{Page}}}, \bibinfo {author} {\bibfnamefont
  {K.~M.}\ \bibnamefont {{Smith}}}, \bibinfo {author} {\bibfnamefont {J.~L.}\
  \bibnamefont {{Weiland}}}, \bibinfo {author} {\bibfnamefont {B.}~\bibnamefont
  {{Gold}}}, \bibinfo {author} {\bibfnamefont {N.}~\bibnamefont {{Jarosik}}},
  \bibinfo {author} {\bibfnamefont {A.}~\bibnamefont {{Kogut}}}, \bibinfo
  {author} {\bibfnamefont {M.}~\bibnamefont {{Limon}}}, \bibinfo {author}
  {\bibfnamefont {S.~S.}\ \bibnamefont {{Meyer}}}, \bibinfo {author}
  {\bibfnamefont {G.~S.}\ \bibnamefont {{Tucker}}}, \bibinfo {author}
  {\bibfnamefont {E.}~\bibnamefont {{Wollack}}}, \ and\ \bibinfo {author}
  {\bibfnamefont {E.~L.}\ \bibnamefont {{Wright}}},\ }\bibfield  {title}
  {\enquote {\bibinfo {title} {{Nine-year Wilkinson Microwave Anisotropy Probe
  (WMAP) Observations: Cosmological Parameter Results}},}\ }\href {\doibase
  10.1088/0067-0049/208/2/19} {\bibfield  {journal} {\bibinfo  {journal} {The
  Astrophysical Journal Supplement Series}\ }\textbf {\bibinfo {volume}
  {208}},\ \bibinfo {eid} {19} (\bibinfo {year} {2013})},\ \Eprint
  {http://arxiv.org/abs/1212.5226} {arXiv:1212.5226 [astro-ph.CO]} \BibitemShut
  {NoStop}%
\bibitem [{\citenamefont {Bennett}\ \emph {et~al.}(2011)\citenamefont
  {Bennett}, \citenamefont {Hill}, \citenamefont {Hinshaw}, \citenamefont
  {Larson}, \citenamefont {Smith}, \citenamefont {Dunkley}, \citenamefont
  {Gold}, \citenamefont {Halpern}, \citenamefont {Jarosik}, \citenamefont
  {Kogut}, \citenamefont {Komatsu}, \citenamefont {Limon}, \citenamefont
  {Meyer}, \citenamefont {Nolta}, \citenamefont {Odegard}, \citenamefont
  {Page}, \citenamefont {Spergel}, \citenamefont {Tucker}, \citenamefont
  {Weiland}, \citenamefont {Wollack},\ and\ \citenamefont
  {Wright}}]{WMAPanomalies}%
  \BibitemOpen
  \bibfield  {author} {\bibinfo {author} {\bibfnamefont {C.~L.}\ \bibnamefont
  {Bennett}}, \bibinfo {author} {\bibfnamefont {R.~S.}\ \bibnamefont {Hill}},
  \bibinfo {author} {\bibfnamefont {G.}~\bibnamefont {Hinshaw}}, \bibinfo
  {author} {\bibfnamefont {D.}~\bibnamefont {Larson}}, \bibinfo {author}
  {\bibfnamefont {K.~M.}\ \bibnamefont {Smith}}, \bibinfo {author}
  {\bibfnamefont {J.}~\bibnamefont {Dunkley}}, \bibinfo {author} {\bibfnamefont
  {B.}~\bibnamefont {Gold}}, \bibinfo {author} {\bibfnamefont {M.}~\bibnamefont
  {Halpern}}, \bibinfo {author} {\bibfnamefont {N.}~\bibnamefont {Jarosik}},
  \bibinfo {author} {\bibfnamefont {A.}~\bibnamefont {Kogut}}, \bibinfo
  {author} {\bibfnamefont {E.}~\bibnamefont {Komatsu}}, \bibinfo {author}
  {\bibfnamefont {M.}~\bibnamefont {Limon}}, \bibinfo {author} {\bibfnamefont
  {S.~S.}\ \bibnamefont {Meyer}}, \bibinfo {author} {\bibfnamefont {M.~R.}\
  \bibnamefont {Nolta}}, \bibinfo {author} {\bibfnamefont {N.}~\bibnamefont
  {Odegard}}, \bibinfo {author} {\bibfnamefont {L.}~\bibnamefont {Page}},
  \bibinfo {author} {\bibfnamefont {D.~N.}\ \bibnamefont {Spergel}}, \bibinfo
  {author} {\bibfnamefont {G.Â S.}\ \bibnamefont {Tucker}}, \bibinfo {author}
  {\bibfnamefont {J.~L.}\ \bibnamefont {Weiland}}, \bibinfo {author}
  {\bibfnamefont {E.}~\bibnamefont {Wollack}}, \ and\ \bibinfo {author}
  {\bibfnamefont {E.~L.}\ \bibnamefont {Wright}},\ }\bibfield  {title}
  {\enquote {\bibinfo {title} {Seven-year wilkinson microwave anisotropy probe
  (wmap) observations: Are there cosmic microwave background anomalies?}}\
  }\href {http://stacks.iop.org/0067-0049/192/i=2/a=17} {\bibfield  {journal}
  {\bibinfo  {journal} {The Astrophysical Journal Supplement Series}\ }\textbf
  {\bibinfo {volume} {192}},\ \bibinfo {pages} {17} (\bibinfo {year}
  {2011})}\BibitemShut {NoStop}%
\bibitem [{\citenamefont {Ade}\ \emph {et~al.}(2016{\natexlab{a}})\citenamefont
  {Ade} \emph {et~al.}}]{Ade:2015xua}%
  \BibitemOpen
  \bibfield  {author} {\bibinfo {author} {\bibfnamefont {P.~A.~R.}\
  \bibnamefont {Ade}} \emph {et~al.} (\bibinfo {collaboration} {Planck}),\
  }\bibfield  {title} {\enquote {\bibinfo {title} {{Planck 2015 results. XIII.
  Cosmological parameters}},}\ }\href {\doibase 10.1051/0004-6361/201525830}
  {\bibfield  {journal} {\bibinfo  {journal} {Astron. Astrophys.}\ }\textbf
  {\bibinfo {volume} {594}},\ \bibinfo {pages} {A13} (\bibinfo {year}
  {2016}{\natexlab{a}})}\BibitemShut {NoStop}%
\bibitem [{\citenamefont {Ade}\ \emph {et~al.}(2016{\natexlab{b}})\citenamefont
  {Ade} \emph {et~al.}}]{Ade:2015lrj}%
  \BibitemOpen
  \bibfield  {author} {\bibinfo {author} {\bibfnamefont {P.~A.~R.}\
  \bibnamefont {Ade}} \emph {et~al.} (\bibinfo {collaboration} {Planck}),\
  }\bibfield  {title} {\enquote {\bibinfo {title} {{Planck 2015 results. XX.
  Constraints on inflation}},}\ }\href {\doibase 10.1051/0004-6361/201525898}
  {\bibfield  {journal} {\bibinfo  {journal} {Astron. Astrophys.}\ }\textbf
  {\bibinfo {volume} {594}},\ \bibinfo {pages} {A20} (\bibinfo {year}
  {2016}{\natexlab{b}})}\BibitemShut {NoStop}%
\bibitem [{\citenamefont {Ade}\ \emph {et~al.}(2016{\natexlab{c}})\citenamefont
  {Ade} \emph {et~al.}}]{Array:2015xqh}%
  \BibitemOpen
  \bibfield  {author} {\bibinfo {author} {\bibfnamefont {P.~A.~R.}\
  \bibnamefont {Ade}} \emph {et~al.} (\bibinfo {collaboration} {BICEP2, Keck
  Array}),\ }\bibfield  {title} {\enquote {\bibinfo {title} {{Improved
  Constraints on Cosmology and Foregrounds from BICEP2 and Keck Array Cosmic
  Microwave Background Data with Inclusion of 95 GHz Band}},}\ }\href {\doibase
  10.1103/PhysRevLett.116.031302} {\bibfield  {journal} {\bibinfo  {journal}
  {Phys. Rev. Lett.}\ }\textbf {\bibinfo {volume} {116}},\ \bibinfo {pages}
  {031302} (\bibinfo {year} {2016}{\natexlab{c}})}\BibitemShut {NoStop}%
\bibitem [{\citenamefont {Ade}\ \emph {et~al.}(2016{\natexlab{d}})\citenamefont
  {Ade} \emph {et~al.}}]{Ade:2015hxq}%
  \BibitemOpen
  \bibfield  {author} {\bibinfo {author} {\bibfnamefont {P.~A.~R.}\
  \bibnamefont {Ade}} \emph {et~al.} (\bibinfo {collaboration} {Planck}),\
  }\bibfield  {title} {\enquote {\bibinfo {title} {{Planck 2015 results. XVI.
  Isotropy and statistics of the CMB}},}\ }\href {\doibase
  10.1051/0004-6361/201526681} {\bibfield  {journal} {\bibinfo  {journal}
  {Astron. Astrophys.}\ }\textbf {\bibinfo {volume} {594}},\ \bibinfo {pages}
  {A16} (\bibinfo {year} {2016}{\natexlab{d}})},\ \Eprint
  {http://arxiv.org/abs/1506.07135} {arXiv:1506.07135 [astro-ph.CO]}
  \BibitemShut {NoStop}%
\bibitem [{\citenamefont {Aylor}\ \emph {et~al.}(2017)\citenamefont {Aylor}
  \emph {et~al.}}]{Aylor:2017haa}%
  \BibitemOpen
  \bibfield  {author} {\bibinfo {author} {\bibfnamefont {K.}~\bibnamefont
  {Aylor}} \emph {et~al.} (\bibinfo {collaboration} {SPT}),\ }\bibfield
  {title} {\enquote {\bibinfo {title} {{A Comparison of Cosmological Parameters
  Determined from CMB Temperature Power Spectra from the South Pole Telescope
  and the Planck Satellite}},}\ }\href {\doibase 10.3847/1538-4357/aa947b}
  {\bibfield  {journal} {\bibinfo  {journal} {Astrophys. J.}\ }\textbf
  {\bibinfo {volume} {850}},\ \bibinfo {pages} {101} (\bibinfo {year}
  {2017})},\ \Eprint {http://arxiv.org/abs/1706.10286} {arXiv:1706.10286
  [astro-ph.CO]} \BibitemShut {NoStop}%
\bibitem [{\citenamefont {Akrami}\ \emph
  {et~al.}(2018{\natexlab{a}})\citenamefont {Akrami} \emph
  {et~al.}}]{Akrami:2018vks}%
  \BibitemOpen
  \bibfield  {author} {\bibinfo {author} {\bibfnamefont {Y.}~\bibnamefont
  {Akrami}} \emph {et~al.} (\bibinfo {collaboration} {Planck}),\ }\bibfield
  {title} {\enquote {\bibinfo {title} {{Planck 2018 results. I. Overview and
  the cosmological legacy of Planck}},}\ }\href@noop {} {\  (\bibinfo {year}
  {2018}{\natexlab{a}})},\ \Eprint {http://arxiv.org/abs/1807.06205}
  {arXiv:1807.06205 [astro-ph.CO]} \BibitemShut {NoStop}%
\bibitem [{\citenamefont {Aghanim}\ \emph {et~al.}(2018)\citenamefont {Aghanim}
  \emph {et~al.}}]{Aghanim:2018eyx}%
  \BibitemOpen
  \bibfield  {author} {\bibinfo {author} {\bibfnamefont {N.}~\bibnamefont
  {Aghanim}} \emph {et~al.} (\bibinfo {collaboration} {Planck}),\ }\bibfield
  {title} {\enquote {\bibinfo {title} {{Planck 2018 results. VI. Cosmological
  parameters}},}\ }\href@noop {} {\  (\bibinfo {year} {2018})},\ \Eprint
  {http://arxiv.org/abs/1807.06209} {arXiv:1807.06209 [astro-ph.CO]}
  \BibitemShut {NoStop}%
\bibitem [{\citenamefont {Akrami}\ \emph
  {et~al.}(2018{\natexlab{b}})\citenamefont {Akrami} \emph
  {et~al.}}]{Akrami:2018odb}%
  \BibitemOpen
  \bibfield  {author} {\bibinfo {author} {\bibfnamefont {Y.}~\bibnamefont
  {Akrami}} \emph {et~al.} (\bibinfo {collaboration} {Planck}),\ }\bibfield
  {title} {\enquote {\bibinfo {title} {{Planck 2018 results. X. Constraints on
  inflation}},}\ }\href@noop {} {\  (\bibinfo {year} {2018}{\natexlab{b}})},\
  \Eprint {http://arxiv.org/abs/1807.06211} {arXiv:1807.06211 [astro-ph.CO]}
  \BibitemShut {NoStop}%
\bibitem [{\citenamefont {Akrami}\ \emph {et~al.}(2019)\citenamefont {Akrami}
  \emph {et~al.}}]{Akrami:2019bkn}%
  \BibitemOpen
  \bibfield  {author} {\bibinfo {author} {\bibfnamefont {Y.}~\bibnamefont
  {Akrami}} \emph {et~al.} (\bibinfo {collaboration} {Planck}),\ }\bibfield
  {title} {\enquote {\bibinfo {title} {{Planck 2018 results. VII. Isotropy and
  Statistics of the CMB}},}\ }\href@noop {} {\  (\bibinfo {year} {2019})},\
  \Eprint {http://arxiv.org/abs/1906.02552} {arXiv:1906.02552 [astro-ph.CO]}
  \BibitemShut {NoStop}%
\bibitem [{\citenamefont {Kadota}\ \emph {et~al.}(2005)\citenamefont {Kadota},
  \citenamefont {Dodelson}, \citenamefont {Hu},\ and\ \citenamefont
  {Stewart}}]{Kadota:2005hv}%
  \BibitemOpen
  \bibfield  {author} {\bibinfo {author} {\bibfnamefont {Kenji}\ \bibnamefont
  {Kadota}}, \bibinfo {author} {\bibfnamefont {Scott}\ \bibnamefont
  {Dodelson}}, \bibinfo {author} {\bibfnamefont {Wayne}\ \bibnamefont {Hu}}, \
  and\ \bibinfo {author} {\bibfnamefont {Ewan~D.}\ \bibnamefont {Stewart}},\
  }\bibfield  {title} {\enquote {\bibinfo {title} {{Precision of inflaton
  potential reconstruction from CMB using the general slow-roll
  approximation}},}\ }\href {\doibase 10.1103/PhysRevD.72.023510} {\bibfield
  {journal} {\bibinfo  {journal} {Phys. Rev.}\ }\textbf {\bibinfo {volume}
  {D72}},\ \bibinfo {pages} {023510} (\bibinfo {year} {2005})}\BibitemShut
  {NoStop}%
\bibitem [{\citenamefont {Baumann}(2011)}]{Baumann:2009ds}%
  \BibitemOpen
  \bibfield  {author} {\bibinfo {author} {\bibfnamefont {Daniel}\ \bibnamefont
  {Baumann}},\ }\bibfield  {title} {\enquote {\bibinfo {title} {{Inflation}},}\
  }in\ \href {\doibase 10.1142/9789814327183_0010} {\emph {\bibinfo {booktitle}
  {{Physics of the large and the small, TASI 09}}}}\ (\bibinfo {year} {2011})\
  pp.\ \bibinfo {pages} {523--686},\ \Eprint {http://arxiv.org/abs/0907.5424}
  {arXiv:0907.5424 [hep-th]} \BibitemShut {NoStop}%
\bibitem [{\citenamefont {Kamionkowski}\ and\ \citenamefont
  {Kovetz}(2016)}]{Kamionkowski:2015yta}%
  \BibitemOpen
  \bibfield  {author} {\bibinfo {author} {\bibfnamefont {Marc}\ \bibnamefont
  {Kamionkowski}}\ and\ \bibinfo {author} {\bibfnamefont {Ely~D.}\ \bibnamefont
  {Kovetz}},\ }\bibfield  {title} {\enquote {\bibinfo {title} {{The Quest for B
  Modes from Inflationary Gravitational Waves}},}\ }\href {\doibase
  10.1146/annurev-astro-081915-023433} {\bibfield  {journal} {\bibinfo
  {journal} {Ann. Rev. Astron. Astrophys.}\ }\textbf {\bibinfo {volume} {54}},\
  \bibinfo {pages} {227--269} (\bibinfo {year} {2016})}\BibitemShut {NoStop}%
\bibitem [{\citenamefont {Banks}\ and\ \citenamefont
  {Fischler}(2018)}]{Banks:2018ypk}%
  \BibitemOpen
  \bibfield  {author} {\bibinfo {author} {\bibfnamefont {Tom}\ \bibnamefont
  {Banks}}\ and\ \bibinfo {author} {\bibfnamefont {W.}~\bibnamefont
  {Fischler}},\ }\bibfield  {title} {\enquote {\bibinfo {title} {{The
  holographic spacetime model of cosmology}},}\ }\href {\doibase
  10.1142/S0218271818460057} {\bibfield  {journal} {\bibinfo  {journal} {Int.
  J. Mod. Phys.}\ }\textbf {\bibinfo {volume} {D27}},\ \bibinfo {pages}
  {1846005} (\bibinfo {year} {2018})},\ \Eprint
  {http://arxiv.org/abs/1806.01749} {arXiv:1806.01749 [hep-th]} \BibitemShut
  {NoStop}%
\bibitem [{\citenamefont {Hogan}(2019)}]{PhysRevD.99.063531}%
  \BibitemOpen
  \bibfield  {author} {\bibinfo {author} {\bibfnamefont {Craig}\ \bibnamefont
  {Hogan}},\ }\bibfield  {title} {\enquote {\bibinfo {title} {Nonlocal
  entanglement and directional correlations of primordial perturbations on the
  inflationary horizon},}\ }\href {\doibase 10.1103/PhysRevD.99.063531}
  {\bibfield  {journal} {\bibinfo  {journal} {Phys. Rev. D}\ }\textbf {\bibinfo
  {volume} {99}},\ \bibinfo {pages} {063531} (\bibinfo {year}
  {2019})}\BibitemShut {NoStop}%
\bibitem [{\citenamefont {'t~Hooft}(2016{\natexlab{a}})}]{Hooft:2016cpw}%
  \BibitemOpen
  \bibfield  {author} {\bibinfo {author} {\bibfnamefont {Gerard}\ \bibnamefont
  {'t~Hooft}},\ }\bibfield  {title} {\enquote {\bibinfo {title} {{The Quantum
  Black Hole as a Hydrogen Atom: Microstates Without Strings Attached}},}\
  }\href@noop {} {\  (\bibinfo {year} {2016}{\natexlab{a}})},\ \Eprint
  {http://arxiv.org/abs/1605.05119} {arXiv:1605.05119 [gr-qc]} \BibitemShut
  {NoStop}%
\bibitem [{\citenamefont {'t~Hooft}(2016{\natexlab{b}})}]{Hooft:2016itl}%
  \BibitemOpen
  \bibfield  {author} {\bibinfo {author} {\bibfnamefont {Gerard}\ \bibnamefont
  {'t~Hooft}},\ }\bibfield  {title} {\enquote {\bibinfo {title} {{Black hole
  unitarity and antipodal entanglement}},}\ }\href {\doibase
  10.1007/s10701-016-0014-y} {\bibfield  {journal} {\bibinfo  {journal} {Found.
  Phys.}\ }\textbf {\bibinfo {volume} {46}},\ \bibinfo {pages} {1185--1198}
  (\bibinfo {year} {2016}{\natexlab{b}})},\ \Eprint
  {http://arxiv.org/abs/1601.03447} {arXiv:1601.03447 [gr-qc]} \BibitemShut
  {NoStop}%
\bibitem [{\citenamefont {'t~Hooft}(2018)}]{Hooft2018}%
  \BibitemOpen
  \bibfield  {author} {\bibinfo {author} {\bibfnamefont {Gerard}\ \bibnamefont
  {'t~Hooft}},\ }\bibfield  {title} {\enquote {\bibinfo {title} {Virtual black
  holes and space--time structure},}\ }\href {\doibase
  10.1007/s10701-017-0133-0} {\bibfield  {journal} {\bibinfo  {journal}
  {Foundations of Physics}\ }\textbf {\bibinfo {volume} {48}},\ \bibinfo
  {pages} {1134--1149} (\bibinfo {year} {2018})}\BibitemShut {NoStop}%
\bibitem [{\citenamefont {Giddings}(2019{\natexlab{a}})}]{Giddings:2018koz}%
  \BibitemOpen
  \bibfield  {author} {\bibinfo {author} {\bibfnamefont {Steven~B.}\
  \bibnamefont {Giddings}},\ }\bibfield  {title} {\enquote {\bibinfo {title}
  {{Quantum-first gravity}},}\ }\href {\doibase 10.1007/s10701-019-00239-1}
  {\bibfield  {journal} {\bibinfo  {journal} {Found. Phys.}\ }\textbf {\bibinfo
  {volume} {49}},\ \bibinfo {pages} {177--190} (\bibinfo {year}
  {2019}{\natexlab{a}})},\ \Eprint {http://arxiv.org/abs/1803.04973}
  {arXiv:1803.04973 [hep-th]} \BibitemShut {NoStop}%
\bibitem [{\citenamefont {Giddings}(2019{\natexlab{b}})}]{Giddings:2019vvj}%
  \BibitemOpen
  \bibfield  {author} {\bibinfo {author} {\bibfnamefont {Steven~B.}\
  \bibnamefont {Giddings}},\ }\bibfield  {title} {\enquote {\bibinfo {title}
  {{Black holes in the quantum universe}},}\ }\bibfield  {booktitle} {\emph
  {\bibinfo {booktitle} {{Proceedings, Topological avatars of new physics:
  London, United Kingdom, March 4-5, 2019}}},\ }\href {\doibase
  10.1098/rsta.2019.0029} {\bibfield  {journal} {\bibinfo  {journal} {Phil.
  Trans. Roy. Soc. Lond.}\ }\textbf {\bibinfo {volume} {A377}},\ \bibinfo
  {pages} {20190029} (\bibinfo {year} {2019}{\natexlab{b}})},\ \Eprint
  {http://arxiv.org/abs/1905.08807} {arXiv:1905.08807 [hep-th]} \BibitemShut
  {NoStop}%
\bibitem [{\citenamefont {Hogan}(2002{\natexlab{a}})}]{Hogan:2002xs}%
  \BibitemOpen
  \bibfield  {author} {\bibinfo {author} {\bibfnamefont {Craig~J.}\
  \bibnamefont {Hogan}},\ }\bibfield  {title} {\enquote {\bibinfo {title}
  {{Holographic discreteness of inflationary perturbations}},}\ }\href
  {\doibase 10.1103/PhysRevD.66.023521} {\bibfield  {journal} {\bibinfo
  {journal} {Phys. Rev.}\ }\textbf {\bibinfo {volume} {D66}},\ \bibinfo {pages}
  {023521} (\bibinfo {year} {2002}{\natexlab{a}})}\BibitemShut {NoStop}%
\bibitem [{\citenamefont {Hogan}(2004)}]{Hogan:2003mq}%
  \BibitemOpen
  \bibfield  {author} {\bibinfo {author} {\bibfnamefont {Craig~J.}\
  \bibnamefont {Hogan}},\ }\bibfield  {title} {\enquote {\bibinfo {title}
  {{Discrete spectrum of inflationary fluctuations}},}\ }\href {\doibase
  10.1103/PhysRevD.70.083521} {\bibfield  {journal} {\bibinfo  {journal} {Phys.
  Rev.}\ }\textbf {\bibinfo {volume} {D70}},\ \bibinfo {pages} {083521}
  (\bibinfo {year} {2004})}\BibitemShut {NoStop}%
\bibitem [{\citenamefont {Hogan}(2002{\natexlab{b}})}]{10.2307/27857721}%
  \BibitemOpen
  \bibfield  {author} {\bibinfo {author} {\bibfnamefont {Craig~J.}\
  \bibnamefont {Hogan}},\ }\bibfield  {title} {\enquote {\bibinfo {title}
  {Observing the beginning of time: New maps of the cosmic background radiation
  may display evidence of the quantum origin of space and time},}\ }\href
  {http://www.jstor.org/stable/27857721} {\bibfield  {journal} {\bibinfo
  {journal} {American Scientist}\ }\textbf {\bibinfo {volume} {90}},\ \bibinfo
  {pages} {420--427} (\bibinfo {year} {2002}{\natexlab{b}})}\BibitemShut
  {NoStop}%
\bibitem [{\citenamefont {Scardigli}\ \emph {et~al.}(2011)\citenamefont
  {Scardigli}, \citenamefont {Gruber},\ and\ \citenamefont
  {Chen}}]{Scardigli:2010gm}%
  \BibitemOpen
  \bibfield  {author} {\bibinfo {author} {\bibfnamefont {Fabio}\ \bibnamefont
  {Scardigli}}, \bibinfo {author} {\bibfnamefont {Christine}\ \bibnamefont
  {Gruber}}, \ and\ \bibinfo {author} {\bibfnamefont {Pisin}\ \bibnamefont
  {Chen}},\ }\bibfield  {title} {\enquote {\bibinfo {title} {{Black Hole
  Remnants in the Early Universe}},}\ }\href {\doibase
  10.1103/PhysRevD.83.063507} {\bibfield  {journal} {\bibinfo  {journal} {Phys.
  Rev.}\ }\textbf {\bibinfo {volume} {D83}},\ \bibinfo {pages} {063507}
  (\bibinfo {year} {2011})},\ \Eprint {http://arxiv.org/abs/1009.0882}
  {arXiv:1009.0882 [gr-qc]} \BibitemShut {NoStop}%
\bibitem [{\citenamefont {Bardeen}(1980)}]{PhysRevD.22.1882}%
  \BibitemOpen
  \bibfield  {author} {\bibinfo {author} {\bibfnamefont {James~M.}\
  \bibnamefont {Bardeen}},\ }\bibfield  {title} {\enquote {\bibinfo {title}
  {Gauge-invariant cosmological perturbations},}\ }\href {\doibase
  10.1103/PhysRevD.22.1882} {\bibfield  {journal} {\bibinfo  {journal} {Phys.
  Rev. D}\ }\textbf {\bibinfo {volume} {22}},\ \bibinfo {pages} {1882--1905}
  (\bibinfo {year} {1980})}\BibitemShut {NoStop}%
\bibitem [{\citenamefont {{Sachs}}\ and\ \citenamefont
  {{Wolfe}}(1967)}]{1967ApJ...147...73S}%
  \BibitemOpen
  \bibfield  {author} {\bibinfo {author} {\bibfnamefont {R.~K.}\ \bibnamefont
  {{Sachs}}}\ and\ \bibinfo {author} {\bibfnamefont {A.~M.}\ \bibnamefont
  {{Wolfe}}},\ }\bibfield  {title} {\enquote {\bibinfo {title} {{Perturbations
  of a Cosmological Model and Angular Variations of the Microwave
  Background}},}\ }\href {\doibase 10.1086/148982} {\bibfield  {journal}
  {\bibinfo  {journal} {\apj}\ }\textbf {\bibinfo {volume} {147}},\ \bibinfo
  {pages} {73} (\bibinfo {year} {1967})}\BibitemShut {NoStop}%
\bibitem [{\citenamefont {Hu}\ and\ \citenamefont
  {Dodelson}(2002)}]{Hu:2001bc}%
  \BibitemOpen
  \bibfield  {author} {\bibinfo {author} {\bibfnamefont {Wayne}\ \bibnamefont
  {Hu}}\ and\ \bibinfo {author} {\bibfnamefont {Scott}\ \bibnamefont
  {Dodelson}},\ }\bibfield  {title} {\enquote {\bibinfo {title} {{Cosmic
  microwave background anisotropies}},}\ }\href {\doibase
  10.1146/annurev.astro.40.060401.093926} {\bibfield  {journal} {\bibinfo
  {journal} {Ann. Rev. Astron. Astrophys.}\ }\textbf {\bibinfo {volume} {40}},\
  \bibinfo {pages} {171--216} (\bibinfo {year} {2002})},\ \Eprint
  {http://arxiv.org/abs/astro-ph/0110414} {arXiv:astro-ph/0110414 [astro-ph]}
  \BibitemShut {NoStop}%
\bibitem [{\citenamefont {Wright}(2003)}]{Wright:2003ig}%
  \BibitemOpen
  \bibfield  {author} {\bibinfo {author} {\bibfnamefont {E.~L.}\ \bibnamefont
  {Wright}},\ }\bibfield  {title} {\enquote {\bibinfo {title} {{Theoretical
  overview of cosmic microwave background anisotropy}},}\ }in\ \href@noop {}
  {\emph {\bibinfo {booktitle} {{Measuring and modeling the universe.
  Proceedings, Symposium, Pasadena, USA, November 17-22, 2002}}}}\ (\bibinfo
  {year} {2003})\ pp.\ \bibinfo {pages} {291--308},\ \Eprint
  {http://arxiv.org/abs/astro-ph/0305591} {arXiv:astro-ph/0305591 [astro-ph]}
  \BibitemShut {NoStop}%
\bibitem [{\citenamefont {{Yadav}}\ and\ \citenamefont
  {{Wandelt}}(2005)}]{2005PhRvD..71l3004Y}%
  \BibitemOpen
  \bibfield  {author} {\bibinfo {author} {\bibfnamefont {Amit~P.}\ \bibnamefont
  {{Yadav}}}\ and\ \bibinfo {author} {\bibfnamefont {Benjamin~D.}\ \bibnamefont
  {{Wandelt}}},\ }\bibfield  {title} {\enquote {\bibinfo {title} {{CMB
  tomography: Reconstruction of adiabatic primordial scalar potential using
  temperature and polarization maps}},}\ }\href {\doibase
  10.1103/PhysRevD.71.123004} {\bibfield  {journal} {\bibinfo  {journal}
  {\prd}\ }\textbf {\bibinfo {volume} {71}},\ \bibinfo {eid} {123004} (\bibinfo
  {year} {2005})},\ \Eprint {http://arxiv.org/abs/astro-ph/0505386}
  {arXiv:astro-ph/0505386 [astro-ph]} \BibitemShut {NoStop}%
\bibitem [{\citenamefont {Dorn}\ \emph {et~al.}(2015)\citenamefont {Dorn},
  \citenamefont {Greiner},\ and\ \citenamefont {Enßlin}}]{Dorn:2014rua}%
  \BibitemOpen
  \bibfield  {author} {\bibinfo {author} {\bibfnamefont {Sebastian}\
  \bibnamefont {Dorn}}, \bibinfo {author} {\bibfnamefont {Maksim}\ \bibnamefont
  {Greiner}}, \ and\ \bibinfo {author} {\bibfnamefont {Torsten~A.}\
  \bibnamefont {Enßlin}},\ }\bibfield  {title} {\enquote {\bibinfo {title}
  {{All-sky reconstruction of the primordial scalar potential from WMAP
  temperature data}},}\ }\href {\doibase 10.1088/1475-7516/2015/02/041}
  {\bibfield  {journal} {\bibinfo  {journal} {JCAP}\ }\textbf {\bibinfo
  {volume} {1502}},\ \bibinfo {pages} {041} (\bibinfo {year} {2015})},\ \Eprint
  {http://arxiv.org/abs/1412.8315} {arXiv:1412.8315 [astro-ph.CO]} \BibitemShut
  {NoStop}%
\bibitem [{\citenamefont {Schwarz}\ \emph {et~al.}(2016)\citenamefont
  {Schwarz}, \citenamefont {Copi}, \citenamefont {Huterer},\ and\ \citenamefont
  {Starkman}}]{Schwarz:2015cma}%
  \BibitemOpen
  \bibfield  {author} {\bibinfo {author} {\bibfnamefont {Dominik~J.}\
  \bibnamefont {Schwarz}}, \bibinfo {author} {\bibfnamefont {Craig~J.}\
  \bibnamefont {Copi}}, \bibinfo {author} {\bibfnamefont {Dragan}\ \bibnamefont
  {Huterer}}, \ and\ \bibinfo {author} {\bibfnamefont {Glenn~D.}\ \bibnamefont
  {Starkman}},\ }\bibfield  {title} {\enquote {\bibinfo {title} {{CMB Anomalies
  after Planck}},}\ }\href {\doibase 10.1088/0264-9381/33/18/184001} {\bibfield
   {journal} {\bibinfo  {journal} {Class. Quant. Grav.}\ }\textbf {\bibinfo
  {volume} {33}},\ \bibinfo {pages} {184001} (\bibinfo {year}
  {2016})}\BibitemShut {NoStop}%
\bibitem [{\citenamefont {Hinshaw}\ \emph {et~al.}(1996)\citenamefont
  {Hinshaw}, \citenamefont {Banday}, \citenamefont {Bennett}, \citenamefont
  {G{\'{o}}rski}, \citenamefont {Kogut}, \citenamefont {Lineweaver},
  \citenamefont {Smoot},\ and\ \citenamefont {Wright}}]{Hinshaw_1996}%
  \BibitemOpen
  \bibfield  {author} {\bibinfo {author} {\bibfnamefont {G.}~\bibnamefont
  {Hinshaw}}, \bibinfo {author} {\bibfnamefont {A.~J.}\ \bibnamefont {Banday}},
  \bibinfo {author} {\bibfnamefont {C.~L.}\ \bibnamefont {Bennett}}, \bibinfo
  {author} {\bibfnamefont {K.~M.}\ \bibnamefont {G{\'{o}}rski}}, \bibinfo
  {author} {\bibfnamefont {A.}~\bibnamefont {Kogut}}, \bibinfo {author}
  {\bibfnamefont {C.~H.}\ \bibnamefont {Lineweaver}}, \bibinfo {author}
  {\bibfnamefont {G.~F.}\ \bibnamefont {Smoot}}, \ and\ \bibinfo {author}
  {\bibfnamefont {E.~L.}\ \bibnamefont {Wright}},\ }\bibfield  {title}
  {\enquote {\bibinfo {title} {Two-point correlations in the
  [{ITAL}]{COBE}[/{ITAL}] {DMR} four-year anisotropy maps},}\ }\href {\doibase
  10.1086/310076} {\bibfield  {journal} {\bibinfo  {journal} {The Astrophysical
  Journal}\ }\textbf {\bibinfo {volume} {464}},\ \bibinfo {pages} {L25--L28}
  (\bibinfo {year} {1996})}\BibitemShut {NoStop}%
\bibitem [{\citenamefont {{Aluri}}\ \emph {et~al.}(2017)\citenamefont
  {{Aluri}}, \citenamefont {{Ralston}},\ and\ \citenamefont
  {{Weltman}}}]{2017MNRAS.472.2410A}%
  \BibitemOpen
  \bibfield  {author} {\bibinfo {author} {\bibfnamefont {Pavan~K.}\
  \bibnamefont {{Aluri}}}, \bibinfo {author} {\bibfnamefont {John~P.}\
  \bibnamefont {{Ralston}}}, \ and\ \bibinfo {author} {\bibfnamefont {Amanda}\
  \bibnamefont {{Weltman}}},\ }\bibfield  {title} {\enquote {\bibinfo {title}
  {{Alignments of parity even/odd-only multipoles in CMB}},}\ }\href {\doibase
  10.1093/mnras/stx2112} {\bibfield  {journal} {\bibinfo  {journal} {MNRAS}\
  }\textbf {\bibinfo {volume} {472}},\ \bibinfo {pages} {2410--2421} (\bibinfo
  {year} {2017})},\ \Eprint {http://arxiv.org/abs/1703.07070} {arXiv:1703.07070
  [astro-ph.CO]} \BibitemShut {NoStop}%
\bibitem [{\citenamefont {{Aluri}}\ and\ \citenamefont
  {{Jain}}(2012)}]{2012MNRAS.419.3378A}%
  \BibitemOpen
  \bibfield  {author} {\bibinfo {author} {\bibfnamefont {Pavan~K.}\
  \bibnamefont {{Aluri}}}\ and\ \bibinfo {author} {\bibfnamefont {Pankaj}\
  \bibnamefont {{Jain}}},\ }\bibfield  {title} {\enquote {\bibinfo {title}
  {{Parity asymmetry in the CMBR temperature power spectrum}},}\ }\href
  {\doibase 10.1111/j.1365-2966.2011.19981.x} {\bibfield  {journal} {\bibinfo
  {journal} {MNRAS}\ }\textbf {\bibinfo {volume} {419}},\ \bibinfo {pages}
  {3378--3392} (\bibinfo {year} {2012})},\ \Eprint
  {http://arxiv.org/abs/1108.5894} {arXiv:1108.5894 [astro-ph.CO]} \BibitemShut
  {NoStop}%
\bibitem [{\citenamefont {{Copi}}\ \emph {et~al.}(2015)\citenamefont {{Copi}},
  \citenamefont {{Huterer}}, \citenamefont {{Schwarz}},\ and\ \citenamefont
  {{Starkman}}}]{2015MNRAS.449.3458C}%
  \BibitemOpen
  \bibfield  {author} {\bibinfo {author} {\bibfnamefont {Craig~J.}\
  \bibnamefont {{Copi}}}, \bibinfo {author} {\bibfnamefont {Dragan}\
  \bibnamefont {{Huterer}}}, \bibinfo {author} {\bibfnamefont {Dominik~J.}\
  \bibnamefont {{Schwarz}}}, \ and\ \bibinfo {author} {\bibfnamefont
  {Glenn~D.}\ \bibnamefont {{Starkman}}},\ }\bibfield  {title} {\enquote
  {\bibinfo {title} {{Large-scale alignments from WMAP and Planck}},}\ }\href
  {\doibase 10.1093/mnras/stv501} {\bibfield  {journal} {\bibinfo  {journal}
  {MNRAS}\ }\textbf {\bibinfo {volume} {449}},\ \bibinfo {pages} {3458--3470}
  (\bibinfo {year} {2015})},\ \Eprint {http://arxiv.org/abs/1311.4562}
  {arXiv:1311.4562 [astro-ph.CO]} \BibitemShut {NoStop}%
\bibitem [{\citenamefont {Hagimoto}\ \emph {et~al.}(2020)\citenamefont
  {Hagimoto}, \citenamefont {Hogan}, \citenamefont {Lewin},\ and\ \citenamefont
  {Meyer}}]{Hagimoto_2020}%
  \BibitemOpen
  \bibfield  {author} {\bibinfo {author} {\bibfnamefont {Ray}\ \bibnamefont
  {Hagimoto}}, \bibinfo {author} {\bibfnamefont {Craig}\ \bibnamefont {Hogan}},
  \bibinfo {author} {\bibfnamefont {Collin}\ \bibnamefont {Lewin}}, \ and\
  \bibinfo {author} {\bibfnamefont {Stephan~S.}\ \bibnamefont {Meyer}},\
  }\bibfield  {title} {\enquote {\bibinfo {title} {Symmetries of {CMB}
  temperature correlation at large angular separations},}\ }\href {\doibase
  10.3847/2041-8213/ab62a0} {\bibfield  {journal} {\bibinfo  {journal} {The
  Astrophysical Journal}\ }\textbf {\bibinfo {volume} {888}},\ \bibinfo {pages}
  {L29} (\bibinfo {year} {2020})}\BibitemShut {NoStop}%
\bibitem [{\citenamefont {Qin}\ \emph {et~al.}(2018)\citenamefont {Qin},
  \citenamefont {Howlett}, \citenamefont {Staveley-Smith},\ and\ \citenamefont
  {Hong}}]{Qin:2018hue}%
  \BibitemOpen
  \bibfield  {author} {\bibinfo {author} {\bibfnamefont {Fei}\ \bibnamefont
  {Qin}}, \bibinfo {author} {\bibfnamefont {Cullan}\ \bibnamefont {Howlett}},
  \bibinfo {author} {\bibfnamefont {Lister}\ \bibnamefont {Staveley-Smith}}, \
  and\ \bibinfo {author} {\bibfnamefont {Tao}\ \bibnamefont {Hong}},\
  }\bibfield  {title} {\enquote {\bibinfo {title} {{Bulk flow in the combined
  2MTF and 6dFGSv surveys}},}\ }\href {\doibase 10.1093/mnras/sty928}
  {\bibfield  {journal} {\bibinfo  {journal} {Mon. Not. Roy. Astron. Soc.}\
  }\textbf {\bibinfo {volume} {477}},\ \bibinfo {pages} {5150--5166} (\bibinfo
  {year} {2018})},\ \Eprint {http://arxiv.org/abs/1804.05520} {arXiv:1804.05520
  [astro-ph.CO]} \BibitemShut {NoStop}%
\bibitem [{\citenamefont {Hogan}(2012)}]{Hogan:2010zs}%
  \BibitemOpen
  \bibfield  {author} {\bibinfo {author} {\bibfnamefont {Craig~J.}\
  \bibnamefont {Hogan}},\ }\bibfield  {title} {\enquote {\bibinfo {title}
  {{Interferometers as Probes of Planckian Quantum Geometry}},}\ }\href
  {\doibase 10.1103/PhysRevD.85.064007} {\bibfield  {journal} {\bibinfo
  {journal} {Phys. Rev.}\ }\textbf {\bibinfo {volume} {D85}},\ \bibinfo {pages}
  {064007} (\bibinfo {year} {2012})}\BibitemShut {NoStop}%
\bibitem [{\citenamefont {Kwon}\ and\ \citenamefont
  {Hogan}(2016)}]{Kwon:2014yea}%
  \BibitemOpen
  \bibfield  {author} {\bibinfo {author} {\bibfnamefont {Ohkyung}\ \bibnamefont
  {Kwon}}\ and\ \bibinfo {author} {\bibfnamefont {Craig~J.}\ \bibnamefont
  {Hogan}},\ }\bibfield  {title} {\enquote {\bibinfo {title} {{Interferometric
  Tests of Planckian Quantum Geometry Models}},}\ }\href {\doibase
  10.1088/0264-9381/33/10/105004} {\bibfield  {journal} {\bibinfo  {journal}
  {Class. Quant. Grav.}\ }\textbf {\bibinfo {volume} {33}},\ \bibinfo {pages}
  {105004} (\bibinfo {year} {2016})}\BibitemShut {NoStop}%
\bibitem [{\citenamefont {Hogan}\ and\ \citenamefont
  {Kwon}(2017)}]{Hogan:2015kva}%
  \BibitemOpen
  \bibfield  {author} {\bibinfo {author} {\bibfnamefont {Craig~J}\ \bibnamefont
  {Hogan}}\ and\ \bibinfo {author} {\bibfnamefont {Ohkyung}\ \bibnamefont
  {Kwon}},\ }\bibfield  {title} {\enquote {\bibinfo {title} {Statistical
  measures of planck scale signal correlations in interferometers},}\ }\href
  {http://stacks.iop.org/0264-9381/34/i=7/a=075006} {\bibfield  {journal}
  {\bibinfo  {journal} {Classical and Quantum Gravity}\ }\textbf {\bibinfo
  {volume} {34}},\ \bibinfo {pages} {075006} (\bibinfo {year}
  {2017})}\BibitemShut {NoStop}%
\bibitem [{\citenamefont {{Hogan}}(2017)}]{Hogan:2015b}%
  \BibitemOpen
  \bibfield  {author} {\bibinfo {author} {\bibfnamefont {C.~J.}\ \bibnamefont
  {{Hogan}}},\ }\bibfield  {title} {\enquote {\bibinfo {title} {{Exotic
  Rotational Correlations in Quantum Geometry}},}\ }\href@noop {} {\bibfield
  {journal} {\bibinfo  {journal} {Phys. Rev. D}\ }\textbf {\bibinfo {volume}
  {95}},\ \bibinfo {pages} {104050} (\bibinfo {year} {2017})}\BibitemShut
  {NoStop}%
\bibitem [{\citenamefont {{Hogan}}\ \emph {et~al.}(2017)\citenamefont
  {{Hogan}}, \citenamefont {{Kwon}},\ and\ \citenamefont
  {{Richardson}}}]{Hogan:2016}%
  \BibitemOpen
  \bibfield  {author} {\bibinfo {author} {\bibfnamefont {C.~J.}\ \bibnamefont
  {{Hogan}}}, \bibinfo {author} {\bibfnamefont {O.}~\bibnamefont {{Kwon}}}, \
  and\ \bibinfo {author} {\bibfnamefont {J.}~\bibnamefont {{Richardson}}},\
  }\bibfield  {title} {\enquote {\bibinfo {title} {{Statistical Model of Exotic
  Rotational Correlations in Emergent Space-Time}},}\ }\href@noop {} {\bibfield
   {journal} {\bibinfo  {journal} {Class. Quantum Grav.}\ }\textbf {\bibinfo
  {volume} {34}},\ \bibinfo {pages} {135006} (\bibinfo {year}
  {2017})}\BibitemShut {NoStop}%
\bibitem [{\citenamefont {Verlinde}\ and\ \citenamefont
  {Zurek}(2019)}]{Verlinde:2019xfb}%
  \BibitemOpen
  \bibfield  {author} {\bibinfo {author} {\bibfnamefont {Erik~P.}\ \bibnamefont
  {Verlinde}}\ and\ \bibinfo {author} {\bibfnamefont {Kathryn~M.}\ \bibnamefont
  {Zurek}},\ }\bibfield  {title} {\enquote {\bibinfo {title} {{Observational
  Signatures of Quantum Gravity in Interferometers}},}\ }\href@noop {} {\
  (\bibinfo {year} {2019})},\ \Eprint {http://arxiv.org/abs/1902.08207}
  {arXiv:1902.08207 [gr-qc]} \BibitemShut {NoStop}%
\bibitem [{\citenamefont {{Chou}}\ \emph {et~al.}(2017)\citenamefont {{Chou}},
  \citenamefont {{Glass}}, \citenamefont {{Gustafson}}, \citenamefont
  {{Hogan}}, \citenamefont {{Kamai}}, \citenamefont {{Kwon}}, \citenamefont
  {{Lanza}}, \citenamefont {{McCuller}}, \citenamefont {{Meyer}}, \citenamefont
  {{Richardson}}, \citenamefont {{Stoughton}}, \citenamefont {{Tomlin}},\ and\
  \citenamefont {{Weiss}}}]{holoshear}%
  \BibitemOpen
  \bibfield  {author} {\bibinfo {author} {\bibfnamefont {A.}~\bibnamefont
  {{Chou}}}, \bibinfo {author} {\bibfnamefont {H.}~\bibnamefont {{Glass}}},
  \bibinfo {author} {\bibfnamefont {H.~R.}\ \bibnamefont {{Gustafson}}},
  \bibinfo {author} {\bibfnamefont {C.~J.}\ \bibnamefont {{Hogan}}}, \bibinfo
  {author} {\bibfnamefont {B.~L.}\ \bibnamefont {{Kamai}}}, \bibinfo {author}
  {\bibfnamefont {O.}~\bibnamefont {{Kwon}}}, \bibinfo {author} {\bibfnamefont
  {R.}~\bibnamefont {{Lanza}}}, \bibinfo {author} {\bibfnamefont
  {L.}~\bibnamefont {{McCuller}}}, \bibinfo {author} {\bibfnamefont {S.~S.}\
  \bibnamefont {{Meyer}}}, \bibinfo {author} {\bibfnamefont {J.}~\bibnamefont
  {{Richardson}}}, \bibinfo {author} {\bibfnamefont {C.}~\bibnamefont
  {{Stoughton}}}, \bibinfo {author} {\bibfnamefont {R.}~\bibnamefont
  {{Tomlin}}}, \ and\ \bibinfo {author} {\bibfnamefont {R.}~\bibnamefont
  {{Weiss}}} (\bibinfo {collaboration} {Holometer Collaboration}),\ }\bibfield
  {title} {\enquote {\bibinfo {title} {{Interferometric Constraints on Quantum
  Geometrical Shear Noise Correlations}},}\ }\href {\doibase
  10.1088/1361-6382/aa7bd3} {\bibfield  {journal} {\bibinfo  {journal} {Class.
  Quant. Grav.}\ }\textbf {\bibinfo {volume} {34}},\ \bibinfo {pages} {165005}
  (\bibinfo {year} {2017})}\BibitemShut {NoStop}%
\bibitem [{\citenamefont {Wheeler}(1946)}]{10.2307/3301037}%
  \BibitemOpen
  \bibfield  {author} {\bibinfo {author} {\bibfnamefont {John~Archibald}\
  \bibnamefont {Wheeler}},\ }\bibfield  {title} {\enquote {\bibinfo {title}
  {Problems and prospects in elementary particle research},}\ }\href
  {http://www.jstor.org/stable/3301037} {\bibfield  {journal} {\bibinfo
  {journal} {Proceedings of the American Philosophical Society}\ }\textbf
  {\bibinfo {volume} {90}},\ \bibinfo {pages} {36--47} (\bibinfo {year}
  {1946})}\BibitemShut {NoStop}%
\bibitem [{\citenamefont {Zeilinger}(1999)}]{RevModPhys.71.S288}%
  \BibitemOpen
  \bibfield  {author} {\bibinfo {author} {\bibfnamefont {Anton}\ \bibnamefont
  {Zeilinger}},\ }\bibfield  {title} {\enquote {\bibinfo {title} {{Experiment
  and the foundations of quantum physics}},}\ }\href {\doibase
  10.1103/RevModPhys.71.S288} {\bibfield  {journal} {\bibinfo  {journal} {Rev.
  Mod. Phys.}\ }\textbf {\bibinfo {volume} {71}},\ \bibinfo {pages}
  {S288--S297} (\bibinfo {year} {1999})}\BibitemShut {NoStop}%
\bibitem [{\citenamefont {Rovelli}(1991)}]{Rovelli_1991}%
  \BibitemOpen
  \bibfield  {author} {\bibinfo {author} {\bibfnamefont {C}~\bibnamefont
  {Rovelli}},\ }\bibfield  {title} {\enquote {\bibinfo {title} {What is
  observable in classical and quantum gravity?}}\ }\href {\doibase
  10.1088/0264-9381/8/2/011} {\bibfield  {journal} {\bibinfo  {journal}
  {Classical and Quantum Gravity}\ }\textbf {\bibinfo {volume} {8}},\ \bibinfo
  {pages} {297--316} (\bibinfo {year} {1991})}\BibitemShut {NoStop}%
\bibitem [{\citenamefont {Pikovski}\ \emph {et~al.}(2017)\citenamefont
  {Pikovski}, \citenamefont {Zych}, \citenamefont {Costa},\ and\ \citenamefont
  {Brukner}}]{Pikovski_2017}%
  \BibitemOpen
  \bibfield  {author} {\bibinfo {author} {\bibfnamefont {Igor}\ \bibnamefont
  {Pikovski}}, \bibinfo {author} {\bibfnamefont {Magdalena}\ \bibnamefont
  {Zych}}, \bibinfo {author} {\bibfnamefont {Fabio}\ \bibnamefont {Costa}}, \
  and\ \bibinfo {author} {\bibfnamefont {{\v{C}}aslav}\ \bibnamefont
  {Brukner}},\ }\bibfield  {title} {\enquote {\bibinfo {title} {{Time Dilation
  in Quantum Systems and Decoherence}},}\ }\href {\doibase
  10.1088/1367-2630/aa5d92} {\bibfield  {journal} {\bibinfo  {journal} {New
  Journal of Physics}\ }\textbf {\bibinfo {volume} {19}},\ \bibinfo {pages}
  {025011} (\bibinfo {year} {2017})}\BibitemShut {NoStop}%
\bibitem [{\citenamefont {Unruh}\ and\ \citenamefont
  {Wald}(2017)}]{0034-4885-80-9-092002}%
  \BibitemOpen
  \bibfield  {author} {\bibinfo {author} {\bibfnamefont {William~G}\
  \bibnamefont {Unruh}}\ and\ \bibinfo {author} {\bibfnamefont {Robert~M}\
  \bibnamefont {Wald}},\ }\bibfield  {title} {\enquote {\bibinfo {title}
  {Information loss},}\ }\href
  {http://stacks.iop.org/0034-4885/80/i=9/a=092002} {\bibfield  {journal}
  {\bibinfo  {journal} {Reports on Progress in Physics}\ }\textbf {\bibinfo
  {volume} {80}},\ \bibinfo {pages} {092002} (\bibinfo {year}
  {2017})}\BibitemShut {NoStop}%
\bibitem [{\citenamefont {Cohen}\ \emph {et~al.}(1999)\citenamefont {Cohen},
  \citenamefont {Kaplan},\ and\ \citenamefont
  {Nelson}}]{CohenKaplanNelson1999}%
  \BibitemOpen
  \bibfield  {author} {\bibinfo {author} {\bibfnamefont {A.~G.}\ \bibnamefont
  {Cohen}}, \bibinfo {author} {\bibfnamefont {D.~B.}\ \bibnamefont {Kaplan}}, \
  and\ \bibinfo {author} {\bibfnamefont {A.~E.}\ \bibnamefont {Nelson}},\
  }\bibfield  {title} {\enquote {\bibinfo {title} {Effective field theory,
  black holes, and the cosmological constant},}\ }\href@noop {} {\bibfield
  {journal} {\bibinfo  {journal} {Phys. Rev. Lett.}\ }\textbf {\bibinfo
  {volume} {82}},\ \bibinfo {pages} {4971} (\bibinfo {year}
  {1999})}\BibitemShut {NoStop}%
\bibitem [{\citenamefont {Solodukhin}(2011)}]{Solodukhin:2011gn}%
  \BibitemOpen
  \bibfield  {author} {\bibinfo {author} {\bibfnamefont {Sergey~N.}\
  \bibnamefont {Solodukhin}},\ }\bibfield  {title} {\enquote {\bibinfo {title}
  {{Entanglement entropy of black holes}},}\ }\href {\doibase
  10.12942/lrr-2011-8} {\bibfield  {journal} {\bibinfo  {journal} {Living Rev.
  Rel.}\ }\textbf {\bibinfo {volume} {14}},\ \bibinfo {pages} {8} (\bibinfo
  {year} {2011})}\BibitemShut {NoStop}%
\bibitem [{\citenamefont {Carney}\ \emph {et~al.}(2019)\citenamefont {Carney},
  \citenamefont {Stamp},\ and\ \citenamefont {Taylor}}]{Carney_2019}%
  \BibitemOpen
  \bibfield  {author} {\bibinfo {author} {\bibfnamefont {Daniel}\ \bibnamefont
  {Carney}}, \bibinfo {author} {\bibfnamefont {Philip C~E}\ \bibnamefont
  {Stamp}}, \ and\ \bibinfo {author} {\bibfnamefont {Jacob~M}\ \bibnamefont
  {Taylor}},\ }\bibfield  {title} {\enquote {\bibinfo {title} {Tabletop
  experiments for quantum gravity: a user's manual},}\ }\href {\doibase
  10.1088/1361-6382/aaf9ca} {\bibfield  {journal} {\bibinfo  {journal}
  {Classical and Quantum Gravity}\ }\textbf {\bibinfo {volume} {36}},\ \bibinfo
  {pages} {034001} (\bibinfo {year} {2019})}\BibitemShut {NoStop}%
\bibitem [{\citenamefont {Jacobson}(1995)}]{Jacobson1995}%
  \BibitemOpen
  \bibfield  {author} {\bibinfo {author} {\bibfnamefont {T.}~\bibnamefont
  {Jacobson}},\ }\bibfield  {title} {\enquote {\bibinfo {title}
  {{Thermodynamics of Spacetime: The Einstein Equation of State}},}\
  }\href@noop {} {\bibfield  {journal} {\bibinfo  {journal} {Phys. Rev. Lett.}\
  }\textbf {\bibinfo {volume} {75}},\ \bibinfo {pages} {1260} (\bibinfo {year}
  {1995})}\BibitemShut {NoStop}%
\bibitem [{\citenamefont {Jacobson}(2016)}]{Jacobson:2015hqa}%
  \BibitemOpen
  \bibfield  {author} {\bibinfo {author} {\bibfnamefont {Ted}\ \bibnamefont
  {Jacobson}},\ }\bibfield  {title} {\enquote {\bibinfo {title} {{Entanglement
  Equilibrium and the Einstein Equation}},}\ }\href {\doibase
  10.1103/PhysRevLett.116.201101} {\bibfield  {journal} {\bibinfo  {journal}
  {Phys. Rev. Lett.}\ }\textbf {\bibinfo {volume} {116}},\ \bibinfo {pages}
  {201101} (\bibinfo {year} {2016})}\BibitemShut {NoStop}%
\bibitem [{\citenamefont {Verlinde}(2011)}]{Verlinde2011}%
  \BibitemOpen
  \bibfield  {author} {\bibinfo {author} {\bibfnamefont {E.}~\bibnamefont
  {Verlinde}},\ }\bibfield  {title} {\enquote {\bibinfo {title} {{On the Origin
  of Gravity and the Laws of Newton}},}\ }\href@noop {} {\bibfield  {journal}
  {\bibinfo  {journal} {JHEP}\ }\textbf {\bibinfo {volume} {1104}},\ \bibinfo
  {pages} {029} (\bibinfo {year} {2011})}\BibitemShut {NoStop}%
\bibitem [{\citenamefont {Padmanabhan}(2014)}]{Padmanabhan:2013nxa}%
  \BibitemOpen
  \bibfield  {author} {\bibinfo {author} {\bibfnamefont {T.}~\bibnamefont
  {Padmanabhan}},\ }\bibfield  {title} {\enquote {\bibinfo {title} {{General
  Relativity from a Thermodynamic Perspective}},}\ }\href {\doibase
  10.1007/s10714-014-1673-7} {\bibfield  {journal} {\bibinfo  {journal} {Gen.
  Rel. Grav.}\ }\textbf {\bibinfo {volume} {46}},\ \bibinfo {pages} {1673}
  (\bibinfo {year} {2014})}\BibitemShut {NoStop}%
\bibitem [{\citenamefont {'t~Hooft}(1993)}]{tHooft:1993dmi}%
  \BibitemOpen
  \bibfield  {author} {\bibinfo {author} {\bibfnamefont {Gerard}\ \bibnamefont
  {'t~Hooft}},\ }\bibfield  {title} {\enquote {\bibinfo {title} {{Dimensional
  reduction in quantum gravity}},}\ }\bibfield  {booktitle} {\emph {\bibinfo
  {booktitle} {{Conference on Highlights of Particle and Condensed Matter
  Physics (SALAMFEST) Trieste, Italy, March 8-12, 1993}}},\ }\href@noop {}
  {\bibfield  {journal} {\bibinfo  {journal} {Conf. Proc.}\ }\textbf {\bibinfo
  {volume} {C930308}},\ \bibinfo {pages} {284--296} (\bibinfo {year}
  {1993})}\BibitemShut {NoStop}%
\bibitem [{\citenamefont {Susskind}(1995)}]{Susskind1995}%
  \BibitemOpen
  \bibfield  {author} {\bibinfo {author} {\bibfnamefont {L.}~\bibnamefont
  {Susskind}},\ }\bibfield  {title} {\enquote {\bibinfo {title} {The world as a
  hologram},}\ }\href@noop {} {\bibfield  {journal} {\bibinfo  {journal} {J.
  Math. Phys.}\ }\textbf {\bibinfo {volume} {36}},\ \bibinfo {pages} {6377}
  (\bibinfo {year} {1995})}\BibitemShut {NoStop}%
\bibitem [{\citenamefont {Bousso}(2002)}]{Bousso:2002ju}%
  \BibitemOpen
  \bibfield  {author} {\bibinfo {author} {\bibfnamefont {Raphael}\ \bibnamefont
  {Bousso}},\ }\bibfield  {title} {\enquote {\bibinfo {title} {{The Holographic
  principle}},}\ }\href {\doibase 10.1103/RevModPhys.74.825} {\bibfield
  {journal} {\bibinfo  {journal} {Rev.Mod.Phys.}\ }\textbf {\bibinfo {volume}
  {74}},\ \bibinfo {pages} {825--874} (\bibinfo {year} {2002})}\BibitemShut
  {NoStop}%
\bibitem [{\citenamefont {Ryu}\ and\ \citenamefont
  {Takayanagi}(2006{\natexlab{a}})}]{Ryu:2006bv}%
  \BibitemOpen
  \bibfield  {author} {\bibinfo {author} {\bibfnamefont {Shinsei}\ \bibnamefont
  {Ryu}}\ and\ \bibinfo {author} {\bibfnamefont {Tadashi}\ \bibnamefont
  {Takayanagi}},\ }\bibfield  {title} {\enquote {\bibinfo {title} {{Holographic
  derivation of entanglement entropy from AdS/CFT}},}\ }\href {\doibase
  10.1103/PhysRevLett.96.181602} {\bibfield  {journal} {\bibinfo  {journal}
  {Phys. Rev. Lett.}\ }\textbf {\bibinfo {volume} {96}},\ \bibinfo {pages}
  {181602} (\bibinfo {year} {2006}{\natexlab{a}})}\BibitemShut {NoStop}%
\bibitem [{\citenamefont {Ryu}\ and\ \citenamefont
  {Takayanagi}(2006{\natexlab{b}})}]{Ryu:2006ef}%
  \BibitemOpen
  \bibfield  {author} {\bibinfo {author} {\bibfnamefont {Shinsei}\ \bibnamefont
  {Ryu}}\ and\ \bibinfo {author} {\bibfnamefont {Tadashi}\ \bibnamefont
  {Takayanagi}},\ }\bibfield  {title} {\enquote {\bibinfo {title} {{Aspects of
  Holographic Entanglement Entropy}},}\ }\href {\doibase
  10.1088/1126-6708/2006/08/045} {\bibfield  {journal} {\bibinfo  {journal}
  {JHEP}\ }\textbf {\bibinfo {volume} {08}},\ \bibinfo {pages} {045} (\bibinfo
  {year} {2006}{\natexlab{b}})}\BibitemShut {NoStop}%
\bibitem [{\citenamefont {Natsuume}(2015)}]{Natsuume:2014sfa}%
  \BibitemOpen
  \bibfield  {author} {\bibinfo {author} {\bibfnamefont {Makoto}\ \bibnamefont
  {Natsuume}},\ }\bibfield  {title} {\enquote {\bibinfo {title} {{AdS/CFT
  Duality User Guide}},}\ }\href {\doibase 10.1007/978-4-431-55441-7}
  {\bibfield  {journal} {\bibinfo  {journal} {Lect. Notes Phys.}\ }\textbf
  {\bibinfo {volume} {903}},\ \bibinfo {pages} {pp.1--294} (\bibinfo {year}
  {2015})}\BibitemShut {NoStop}%
\bibitem [{\citenamefont {Hawking}\ \emph {et~al.}(2016)\citenamefont
  {Hawking}, \citenamefont {Perry},\ and\ \citenamefont
  {Strominger}}]{Hawking:2016msc}%
  \BibitemOpen
  \bibfield  {author} {\bibinfo {author} {\bibfnamefont {Stephen~W.}\
  \bibnamefont {Hawking}}, \bibinfo {author} {\bibfnamefont {Malcolm~J.}\
  \bibnamefont {Perry}}, \ and\ \bibinfo {author} {\bibfnamefont {Andrew}\
  \bibnamefont {Strominger}},\ }\bibfield  {title} {\enquote {\bibinfo {title}
  {{Soft Hair on Black Holes}},}\ }\href {\doibase
  10.1103/PhysRevLett.116.231301} {\bibfield  {journal} {\bibinfo  {journal}
  {Phys. Rev. Lett.}\ }\textbf {\bibinfo {volume} {116}},\ \bibinfo {pages}
  {231301} (\bibinfo {year} {2016})},\ \Eprint
  {http://arxiv.org/abs/1601.00921} {arXiv:1601.00921 [hep-th]} \BibitemShut
  {NoStop}%
\bibitem [{\citenamefont {Haco}\ \emph {et~al.}(2018)\citenamefont {Haco},
  \citenamefont {Hawking}, \citenamefont {Perry},\ and\ \citenamefont
  {Strominger}}]{Haco:2018ske}%
  \BibitemOpen
  \bibfield  {author} {\bibinfo {author} {\bibfnamefont {Sasha}\ \bibnamefont
  {Haco}}, \bibinfo {author} {\bibfnamefont {Stephen~W.}\ \bibnamefont
  {Hawking}}, \bibinfo {author} {\bibfnamefont {Malcolm~J.}\ \bibnamefont
  {Perry}}, \ and\ \bibinfo {author} {\bibfnamefont {Andrew}\ \bibnamefont
  {Strominger}},\ }\bibfield  {title} {\enquote {\bibinfo {title} {{Black Hole
  Entropy and Soft Hair}},}\ }\href {\doibase 10.1007/JHEP12(2018)098}
  {\bibfield  {journal} {\bibinfo  {journal} {JHEP}\ }\textbf {\bibinfo
  {volume} {12}},\ \bibinfo {pages} {098} (\bibinfo {year} {2018})},\ \Eprint
  {http://arxiv.org/abs/1810.01847} {arXiv:1810.01847 [hep-th]} \BibitemShut
  {NoStop}%
\bibitem [{\citenamefont {Nomura}(2019)}]{Nomura:2019qps}%
  \BibitemOpen
  \bibfield  {author} {\bibinfo {author} {\bibfnamefont {Yasunori}\
  \bibnamefont {Nomura}},\ }\bibfield  {title} {\enquote {\bibinfo {title}
  {{Spacetime and Universal Soft Modes --- Black Holes and Beyond}},}\
  }\href@noop {} {\  (\bibinfo {year} {2019})},\ \Eprint
  {http://arxiv.org/abs/1908.05728} {arXiv:1908.05728 [hep-th]} \BibitemShut
  {NoStop}%
\bibitem [{\citenamefont {Dray}\ and\ \citenamefont
  {'t~Hooft}(1985)}]{DRAY1985173}%
  \BibitemOpen
  \bibfield  {author} {\bibinfo {author} {\bibfnamefont {Tevian}\ \bibnamefont
  {Dray}}\ and\ \bibinfo {author} {\bibfnamefont {Gerard}\ \bibnamefont
  {'t~Hooft}},\ }\bibfield  {title} {\enquote {\bibinfo {title} {The
  gravitational shock wave of a massless particle},}\ }\href {\doibase
  https://doi.org/10.1016/0550-3213(85)90525-5} {\bibfield  {journal} {\bibinfo
   {journal} {Nuclear Physics B}\ }\textbf {\bibinfo {volume} {253}},\ \bibinfo
  {pages} {173 -- 188} (\bibinfo {year} {1985})}\BibitemShut {NoStop}%
\bibitem [{\citenamefont {Aichelburg}\ and\ \citenamefont
  {Sexl}(1971)}]{Aichelburg1971}%
  \BibitemOpen
  \bibfield  {author} {\bibinfo {author} {\bibfnamefont {P.~C.}\ \bibnamefont
  {Aichelburg}}\ and\ \bibinfo {author} {\bibfnamefont {R.~U.}\ \bibnamefont
  {Sexl}},\ }\bibfield  {title} {\enquote {\bibinfo {title} {On the
  gravitational field of a massless particle},}\ }\href {\doibase
  10.1007/BF00758149} {\bibfield  {journal} {\bibinfo  {journal} {General
  Relativity and Gravitation}\ }\textbf {\bibinfo {volume} {2}},\ \bibinfo
  {pages} {303--312} (\bibinfo {year} {1971})}\BibitemShut {NoStop}%
\bibitem [{\citenamefont {Chatwin-Davies}\ \emph {et~al.}(2017)\citenamefont
  {Chatwin-Davies}, \citenamefont {Kempf},\ and\ \citenamefont
  {Martin}}]{Chatwin-Davies:2016byj}%
  \BibitemOpen
  \bibfield  {author} {\bibinfo {author} {\bibfnamefont {Aidan}\ \bibnamefont
  {Chatwin-Davies}}, \bibinfo {author} {\bibfnamefont {Achim}\ \bibnamefont
  {Kempf}}, \ and\ \bibinfo {author} {\bibfnamefont {Robert T.~W.}\
  \bibnamefont {Martin}},\ }\bibfield  {title} {\enquote {\bibinfo {title}
  {{Natural Covariant Planck Scale Cutoffs and the Cosmic Microwave Background
  Spectrum}},}\ }\href {\doibase 10.1103/PhysRevLett.119.031301} {\bibfield
  {journal} {\bibinfo  {journal} {Phys. Rev. Lett.}\ }\textbf {\bibinfo
  {volume} {119}},\ \bibinfo {pages} {031301} (\bibinfo {year} {2017})},\
  \Eprint {http://arxiv.org/abs/1612.06445} {arXiv:1612.06445 [gr-qc]}
  \BibitemShut {NoStop}%
\bibitem [{\citenamefont {Hogan}(2008{\natexlab{a}})}]{Hogan:2007pk}%
  \BibitemOpen
  \bibfield  {author} {\bibinfo {author} {\bibfnamefont {Craig~J.}\
  \bibnamefont {Hogan}},\ }\bibfield  {title} {\enquote {\bibinfo {title}
  {{Measurement of Quantum Fluctuations in Geometry}},}\ }\href {\doibase
  10.1103/PhysRevD.77.104031} {\bibfield  {journal} {\bibinfo  {journal} {Phys.
  Rev.}\ }\textbf {\bibinfo {volume} {D77}},\ \bibinfo {pages} {104031}
  (\bibinfo {year} {2008}{\natexlab{a}})},\ \Eprint
  {http://arxiv.org/abs/0712.3419} {arXiv:0712.3419 [gr-qc]} \BibitemShut
  {NoStop}%
\bibitem [{\citenamefont {Hogan}(2008{\natexlab{b}})}]{Hogan:2008zw}%
  \BibitemOpen
  \bibfield  {author} {\bibinfo {author} {\bibfnamefont {Craig~J.}\
  \bibnamefont {Hogan}},\ }\bibfield  {title} {\enquote {\bibinfo {title}
  {{Indeterminacy of Holographic Quantum Geometry}},}\ }\href {\doibase
  10.1103/PhysRevD.78.087501} {\bibfield  {journal} {\bibinfo  {journal} {Phys.
  Rev.}\ }\textbf {\bibinfo {volume} {D78}},\ \bibinfo {pages} {087501}
  (\bibinfo {year} {2008}{\natexlab{b}})},\ \Eprint
  {http://arxiv.org/abs/0806.0665} {arXiv:0806.0665 [gr-qc]} \BibitemShut
  {NoStop}%
\bibitem [{\citenamefont {Starobinsky}(1980)}]{STAROBINSKY198099}%
  \BibitemOpen
  \bibfield  {author} {\bibinfo {author} {\bibfnamefont {A.A.}\ \bibnamefont
  {Starobinsky}},\ }\bibfield  {title} {\enquote {\bibinfo {title} {A new type
  of isotropic cosmological models without singularity},}\ }\href {\doibase
  https://doi.org/10.1016/0370-2693(80)90670-X} {\bibfield  {journal} {\bibinfo
   {journal} {Physics Letters B}\ }\textbf {\bibinfo {volume} {91}},\ \bibinfo
  {pages} {99 -- 102} (\bibinfo {year} {1980})}\BibitemShut {NoStop}%
\bibitem [{\citenamefont {Polchinski}(1998)}]{polchinski1998string}%
  \BibitemOpen
  \bibfield  {author} {\bibinfo {author} {\bibfnamefont {J.}~\bibnamefont
  {Polchinski}},\ }\href@noop {} {\emph {\bibinfo {title} {String Theory}}}\
  (\bibinfo  {publisher} {Cambridge University Press},\ \bibinfo {year}
  {1998})\BibitemShut {NoStop}%
\bibitem [{\citenamefont {Schwarz}(2000)}]{schwarz}%
  \BibitemOpen
  \bibfield  {author} {\bibinfo {author} {\bibfnamefont {J.~H.}\ \bibnamefont
  {Schwarz}},\ }\href@noop {} {\enquote {\bibinfo {title} {Introduction to
  superstring theory},}\ } (\bibinfo {year} {2000}),\ \Eprint
  {http://arxiv.org/abs/hep-ex/0008017} {arXiv:hep-ex/0008017} \BibitemShut
  {NoStop}%
\bibitem [{\citenamefont {Rovelli}(2004)}]{Rovelli2004}%
  \BibitemOpen
  \bibfield  {author} {\bibinfo {author} {\bibfnamefont {C.}~\bibnamefont
  {Rovelli}},\ }\href@noop {} {\emph {\bibinfo {title} {Quantum Gravity}}}\
  (\bibinfo  {publisher} {Cambridge University Press},\ \bibinfo {year}
  {2004})\BibitemShut {NoStop}%
\bibitem [{\citenamefont {Dirac}(1958)}]{dirac}%
  \BibitemOpen
  \bibfield  {author} {\bibinfo {author} {\bibfnamefont {P.~A.~M.}\
  \bibnamefont {Dirac}},\ }\href@noop {} {\emph {\bibinfo {title} {Quantum
  Mechanics}}}\ (\bibinfo  {publisher} {Oxford University Press},\ \bibinfo
  {address} {Oxford, UK},\ \bibinfo {year} {1958})\BibitemShut {NoStop}%
\bibitem [{\citenamefont {Weinberg}(1996)}]{weinberg1996quantum}%
  \BibitemOpen
  \bibfield  {author} {\bibinfo {author} {\bibfnamefont {S.}~\bibnamefont
  {Weinberg}},\ }\href@noop {} {\emph {\bibinfo {title} {The Quantum Theory of
  Fields}}}\ (\bibinfo  {publisher} {Cambridge University Press},\ \bibinfo
  {year} {1996})\BibitemShut {NoStop}%
\bibitem [{\citenamefont {{Dicke, R.H. and Wittke, J. P.}}(1960)}]{dicke}%
  \BibitemOpen
  \bibfield  {author} {\bibinfo {author} {\bibnamefont {{Dicke, R.H. and
  Wittke, J. P.}}},\ }\href@noop {} {\emph {\bibinfo {title} {{Introduction to
  Quantum Mechanics}}}}\ (\bibinfo  {publisher} {Reading: Addison-Wesley},\
  \bibinfo {year} {1960})\BibitemShut {NoStop}%
\bibitem [{\citenamefont {{Landau, L. D. and Lifshitz, E. M.}}(1977)}]{landau}%
  \BibitemOpen
  \bibfield  {author} {\bibinfo {author} {\bibnamefont {{Landau, L. D. and
  Lifshitz, E. M.}}},\ }\href@noop {} {\emph {\bibinfo {title} {{Quantum
  Mechanics: Non-Relativistic Theory}}}}\ (\bibinfo  {publisher} {Oxford:
  Pergamon},\ \bibinfo {year} {1977})\BibitemShut {NoStop}%
\end{thebibliography}%

\section*{Appendix}

\subsection{Theoretical motivation for  a coherent quantum inflationary horizon}

The main hypothesis of holographic inflation is that the boundary between classical and quantum descriptions of geometry  is a  null surface.  This long-known concept\cite{10.2307/3301037} allows a covariant reduction of  quantum geometrical states to 
connect  consistently with standard  non-dynamical ``spooky'' spacelike quantum relationships among particle states\cite{RevModPhys.71.S288,Rovelli_1991,Pikovski_2017}.  Standard quantum inflation adopts a model  quantum system based on effective field theory and linearized gravity, and does not have this property: its decomposition into comoving waves is motivated by linear evolution equations based on continuous derivatives, which depend on classical locality.  It is useful to summarize some current threads of gravitational theory that motivate the holographic model.

In an explicit quantum model of eternal black holes that includes gravitational back-reaction by emission and absorption of quantum particles\cite{Hooft:2016cpw,Hooft:2016itl,Hooft2018},  the horizon of a black hole is a coherent quantum object, so that a measurement  by an external observer  nonlocally affects its global state.
This hypothesis allows an elegant solution of many paradoxes associated with quantum black holes (e.g., \cite{0034-4885-80-9-092002,CohenKaplanNelson1999,Solodukhin:2011gn,Carney_2019}).  The calculation has been extrapolated to causal diamonds in flat space-time, where it leads to directionally coherent macroscopic fluctuations\cite{Verlinde:2019xfb}.
The main assumption of the current work is that the inflationary horizon has a similar global coherence.

Another powerful line of  argument is based on a derivation of Einstein's field equations\cite{Jacobson1995,Jacobson:2015hqa} from the requirement that any patch of a null surface behaves like a black hole horizon, in the accelerating frame where it is the observer's horizon. 
This result supports the view that classical  space-time and gravitation are emergent, collective  quantum phenomena\cite{Jacobson1995,Jacobson:2015hqa,Verlinde2011,Padmanabhan:2013nxa}, and that
holographic and thermodynamic properties of black hole horizons generalize to universal properties of null surfaces in any  space-time\cite{tHooft:1993dmi,Susskind1995,Bousso:2002ju}.   In holographic inflation, the horizon is taken to be the light cone that bounds the past of an observer at the end of inflation; coherent fluctuations of its nested causal diamonds freeze in as  classical potentials when other comoving locations  pass through it.

Coherent properties of  geometrical states  have also been  extensively studied in  anti-de Sitter space \cite{Ryu:2006bv,Ryu:2006ef,Natsuume:2014sfa}. In this case,  nonlocal spacelike correlations are encoded in a dual theory, that of a conformal field on the lower-dimensional boundary.

 
\subsection{Estimates of coherent fluctuations of causal diamonds and horizons from a semiclassical correspondence principle}

In spite of   considerable theoretical attention recently to the mathematical nature of nonlocalized information in black holes\cite{Hawking:2016msc,Haco:2018ske,Nomura:2019qps}, there is as yet no consensus on the magnitude or physical effects of coherent, large-angle fluctuations of horizons\cite{Giddings:2018koz,Giddings:2019vvj} or causal diamond surfaces \cite{Verlinde:2019xfb}.

Here, we estimate  the magnitude and nonlocal coherence of   physical effects from  Planck scale quantum gravity, using the Bohr correspondence principle for some simple systems. In this view, the quantum system is  a ``sum over histories'', or a superposition of different metrics, without presuming that these states decompose into independent plane-wave perturbations.
Instead, the  macroscopic effect  of a quantum system is required to match its classical behavior, so that  the active gravity of a quantum system of particles can have the same coherence as the particle state. These simple examples show how such coherence  amplifies the effect of Planck scale fluctuations on large scales, and entangles causal structure in different directions.
They indicate that the dominant quantum fluctuations of coherent causal diamonds or horizons of any size $R$ come from the  Planck scale, rather than  fluctuations of a field vacuum modes of wavelength  $R$.


 \begin{figure}
\begin{centering}
\includegraphics[width=\linewidth]{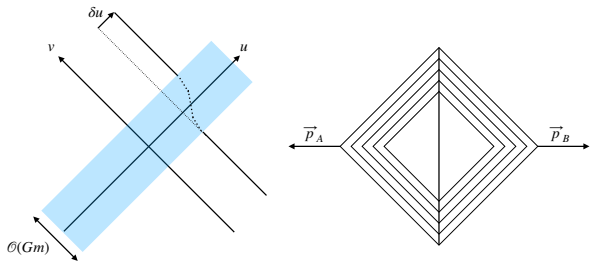}
\par\end{centering}
\protect\caption{ Classical gravitational shocks in nearly-flat space-time from a  point mass $m$, adapted from ref. \cite{DRAY1985173}. 
In null coordinates $u,v$,  the displacement $\delta u$ due to gravitational drag (Eq. \ref{drag}) is shown as a function of $v$ for a point particle  moving in the $u$ direction with momentum $p>>mc$.  The physical displacement is axially symmetric around the particle trajectory and  independent of the impact parameter of the particle, $x_\perp$.   At right, the spatial patterns of  shock waves from  counterpropagating particles are shown. 
 \label{shocks}}
\end{figure}

\subsubsection{Gravitational shock waves of point particles with Planck momentum in flat space-time}

Consider  the classical gravitational effect of a null point particle\cite{Aichelburg1971,DRAY1985173,Hooft:2016cpw,Hooft:2016itl,Hooft2018}, as shown in Fig. (\ref{shocks}).
A localized pointlike momentum impulse $p$ with impact parameter  $x_\perp$ creates an  invariant coherent delocalized transverse displacement   everywhere on a light sheet,
\begin{equation}\label{drag}
\delta u = 4 G p \ln(x_\perp^2).
\end{equation}
This displacement represents a real, physical distortion of causal structure; the motion of matter ``drags'' the space-time along with it.

Now consider the gravitational effect of a  quantum state: a pair of Planck momentum particles emitted in a spherical wave function, like a pair of antipodally propagating gamma rays from annihilations in a PET imager.
For any given emission axis, the shocks create a 
uniform inwards displacement of the surface everywhere on  the equator of the axis  determined by a particle pair, as shown on the right side of Fig. (\ref{shocks}).
Since the axis is indeterminate until the particle wave collapses, the gravitational state must have the same geometrical coherence.
Thus, after a duration $\tau$ the correspondence principle requires coherent nonlocal correlations of displacement $\delta u$ everywhere on a causal diamond surface of size $c\tau$, 
 for consistency of GR with any  axis of the particle pair.
The magnitude of displacement is  independent of the  size of the diamond.

The same correlations between poles and equator apply for a random sequence of impulses in different directions.
The  distortion of causal structure is  additive after each shock wave passes, so the amplitude from many pairs does not cancel, but  adds in quadrature. 
For  Planck-momentum fluctuations  every Planck time (the scale of fluctuations in  models with discrete time at the Planck scale\cite{Chatwin-Davies:2016byj}),
the sum after time $\tau$ is
\begin{equation}
\langle\Delta^2\rangle \sim  t_P/\tau. 
\end{equation}

This estimate agrees with previous estimates based on random walks of world lines, standard quantum uncertainty and wave diffraction
\cite{Hogan:2007pk,Hogan:2008zw,Hogan:2015b}. This set-up  explicitly shows the coherent separation between radial and transverse (angular) degrees of freedom, and the unattenuated macroscopic transverse effects of  Planck scale particles on an arbitrarily large scale.
These coherent  properties of  classical particles should also apply to the gravitational effect of their quantum states; the causal structure needs to inherit the same spooky, transverse correlations.

We have invoked a similar amplitude and transverse coherence  for the  inflationary horizon.
In the inflationary system,  null trajectories curve and in general  acquire transverse components, leading to the angular-domain correlations. 
The flat-space gravitational shock wave  shows  the magnitude of the coherent Planck scale fluctuation that might appear in experimental signals, but does not include the process of freezing on the horizon.
 
\subsubsection{Classical tides from point masses added and subtracted from a black hole}

The back-reaction of quantum point particles entering and leaving a black hole horizon, including antipodal correlations, was studied in refs. \cite{Hooft:2016cpw,Hooft:2016itl,Hooft2018}, with the gravitational effect  based on the gravitational shock wave in the Rindler frame. Here, we use classical correspondence to estimate
 the  tidal distortion of  the shape of the black hole horizon by point particles.
 
A point mass falling  into a black hole of radius $R$
 raises a tide and creates a coherent  distortion of the horizon with a signficant quadrupolar component on a timescale $c/R$.
For addition of  many point  masses in a time $<R/c$,  the 
mean square quadrupole amplitude adds in quadrature. For each mass $m$, $\delta R/R\sim m/M$, so $N$ of them produce $\langle(\delta R/R)^2\rangle \sim N (m/M)^2$.

Quantum  horizons should have similar coherence. Suppose again (e.g., as in \cite{Chatwin-Davies:2016byj}) that there is some form of discreteness in proper time at the Planck scale.
Suppose that each Planck time, a Planck mass particle fluctuates in or out of the hole.
Their mean contributions  average out, but for random orientations,  their  large-angle, low-multipole distortions add in quadrature.
In this case,  $N\sim R/ct_P$, so the 
the variance of the horizon distortion is given by $\langle \Delta^2 \rangle= \langle h^2\rangle= \langle \delta R^2 \rangle /R^2  \sim c t_P/ R$, again the same as the estimate used for holographic horizon.
The states are similarly  nonlocally entangled over a time  $\sim R/c$
like a coherent quantum spin state, creating a  superposition and entanglement of directions.

\subsubsection{Virtual quadrupolar distortion of a black hole necessary to radiate classically at the Hawking rate}

In linear gravity, a source emits  gravitational radiation according to the standard classical quadrupole formula,
 \begin{equation}
h_{wave} = (G/r) \ddot I_{ij},
\end{equation}
where $h_{wave}$ is strain of the wave at distance $r$, and $I_{ij}= \int d^3x \rho x_i x_j$ represents the quadrupole of the matter distribution. For coherent quadrupolar vibrational distortions of a black hole horizon of magnitude
\begin{equation}
h_{horizon}\sim \delta R/R,
\end{equation}
we have approximately
\begin{equation}
\ddot I_{ij}\sim h_{horizon}^2 (R/G).
\end{equation}
Thus,
\begin{equation}
h_{wave} \sim  h_{horizon}^2 (R/r),
\end{equation}
so  in the near-wave zone with $R\sim r$, 
\begin{equation}
h_{wave} \sim  h_{horizon}^2.
\end{equation}
That is,  in the classical system,  gravitational waves are second order compared to the distortions of causal structure that generate them.

In the semiclassical, sum-over-histories view, the virtual metric fluctuations of a quantum  black hole horizon  in the near field are also  much larger than the metric fluctuations  of the Hawking radiation that escapes from the system and carries energy away. 
The  power per area of a classical localized gravitational-wave packet, comparable  in size to wavelength $\sim 2\pi c/\omega$ in all directions, is about $\sim \langle h_{wave}^2\rangle c^3\omega^2/G $; the metric of strain of a single graviton wave packet localized  to  the same volume is
\begin{equation}
\langle h_{wave}^2\rangle \sim (t_P\omega)^2 .
\end{equation}
The typical equivalent classical strain  of Hawking radiation from a black hole of size $R\sim c/\omega$ on the scale $r\sim R$ as it leaves  the vicinity of the hole is $\langle h_{wave}^2\rangle \sim (ct_P/R)^2$;  the  horizons in the quantum superposition vary coherently  with quadrupole amplitude 
\begin{equation}
\langle h_{horizon}^2\rangle \sim (ct_P/R).
\end{equation}
The Hawking radiation can be viewed as an atom-like transition between these states.

During inflation, the equivalent of 
$h_{wave}$ is the metric strain of tensor modes generated by quantum fluctuations\cite{STAROBINSKY198099} on the horizon scale, the zero point oscillations of a field mode  of  frequency $\omega = H$.
In holographic inflation, the  scalar perturbations arise from   coherent Planck scale fluctuations  in the locations (or equivalently, clocks) of world lines on the scale $R= c/H$,  not zero point of field modes
of frequency $H$. 
Thus, as in the classical black hole,  
\begin{equation}
\langle h_{horizon}^2\rangle \sim \langle h_{wave}^2\rangle^{1/2} \sim H t_P
\end{equation}
so scalar perturbations are  much larger than the  tensor distortions, which have
 $\langle h_T^2\rangle = (H t_P)^2/2\pi^2$.

\subsection{Spin-algebra model of coherent quantum causal diamond fluctuations in flat space-time}

The physical consequences of geometrical nonlocality depend on the connection of  quantum mechanics and space-time at a basic level. If geometry and locality are  emergent properties of a quantum system, the quantum-field approximation to the system breaks down, as it omits important entanglements between matter and geometrical degrees of freedom.

This section develops a covariant  model of emergent, coherent quantum causal diamond states in a flat background.  It provides a concretely defined quantum system to  illustrate physical effects of a new form of nonlocal geometrical entanglement,  not included in standard quantized linear gravity.
This model  does not  address quantum dynamics at  the Planck scale, nor is it a substitute for other, arguably more fundamental theories of quantum gravity 
 \cite{polchinski1998string,schwarz,Rovelli2004}.
It is  introduced to reveal measurable physical effects of coherent causal diamond states that differ from effective field theory and linearized gravity:  the large fluctuations introduced by coherence, and new symmetries of transverse or directional correlations.
These effects are found in a regime not easily accessible in other approaches,  on scales much larger than $ct_P$.  

 
 \subsubsection{Dirac light cone function}

Locality on light cones is conventionally defined by the covariant four-dimensional generalization of the one-dimensional Dirac $\delta$-function (ref. \cite{dirac}, \textsection 75): 
\begin{equation}\label{DiracDelta}
\Delta(x) = 2 \delta(x^\mu x_\mu) x_0/|x_0|,
\end{equation}
where $x_\mu = (t, \vec r)$ represents 4-position.
It vanishes at the origin, and is nonvanishing on past and future light cones from the origin.  It is odd in timelike directions and even at spacelike separations, with a
4D point-parity antisymmetry that combines time and space,
\begin{equation}\label{antiDelta}
\Delta(-x) = -  \Delta(x).
\end{equation}
It  has a purely imaginary transform, of the same functional form,
\begin{equation}\label{4DDiractransform}
\tilde\Delta (k) \equiv \int d^4x \Delta(x) e^{ik x} =  4 \pi^2 i \Delta (k).
\end{equation}

\subsubsection{Locality of quantum field states}

The standard model of locality for field states is
based on point localization in a classical space-time background. 
To quantize  fields\cite{dirac,weinberg1996quantum}, the light cone function  is used to write covariant 
 commutation relations for field operators
 $\hat A (x)$,
\begin{equation}\label{fieldcommute}
[\hat A_\mu (x),  \hat A_\nu(x')] =  g_{\mu\nu} \Delta(x-x'),
\end{equation}
where $g_{\mu\nu} $ denotes  a (classical) tensor. 
The explicit geometrical  coefficients on the right hand side  are all classical objects: geometry is not part of the quantum system.
There is an unwritten quantum operator on the right  side which is just the identity operator on a  field state.
Since the only position dependence on the right side comes in  classical functions,
there is no  entanglement of field states with geometry.

For field dynamics, it is necessary to include conjugate momentum or derivative operators.
This leads to  propagating states in the form of  plane waves, with a
 commutator for the transform of  quantized wave modes
\begin{equation}\label{wavecommute}
[\hat A_{\mu}( \vec k),  \hat A_{\nu}( \vec k')] = i g_{\mu\nu} \delta(\vec k -\vec k')/ 4\pi^2 k_0,
\end{equation}
where $g_{\mu\nu} $ again denotes  a (classical) tensor. 
The coefficient depends on the spin or  helicity of the field, and relates internal quantum degrees of freedom to the classical inertial frame. 
In  the case of linearized gravity the propagating quanta are gravitons,
a spin 2 tensor field with a  very small self coupling set by the Planck scale.

 In standard inflation,  perturbations arise from  linearized gravity coupled to quantum fluctuations of field vacuum states. The ``collapse'' of these states into a classical metric occurs coherently in $\vec k$ space,  for each spatially infinite mode.  In holographic or spooky inflation,  these perturbations are subdominant to those from new fundamental geometrical quantum degrees of freedom that underlie holographic, emergent gravity\cite{Banks:2018ypk,PhysRevD.99.063531}.

\subsubsection{Model of coherent  causal diamond fluctuations}

A simple model  of a  geometrical  quantum system can be built from new
 geometrical  operators $\hat {\tau}_\kappa$,  whose fluctuations will ultimately be identified as perturbations in potential.
A contraction with the antisymmetric Levi-Civita 4-tensor $\epsilon_{\kappa\lambda\mu\nu}$ allows us to write a  Lorentz  covariant  generalized rotational commutative algebra  with the same light cone structure
as  the field commutator (Eq. \ref{fieldcommute}):
\begin{equation}\label{qgeo}
[\hat {\tau}_\kappa, \hat {\tau}_\lambda] =  i \epsilon_{\kappa\lambda\mu\nu} \  \hat {\tau}^\mu\  \Delta (x^\nu) \tau_0.
\end{equation}
The  imaginary coefficient in  Eq. (\ref{qgeo}) allows for superposition and entanglement of the geometrical states. 
In Eq. (\ref{qgeo}), geometrical quantum operators on both  sides share the same degrees of freedom, so there  are new nonlocal quantum relations  that cannot be described by Eq. (\ref{fieldcommute}), even with a linearized tensor field to represent quantum gravity.

Eq. (\ref{qgeo})  is not a  fully consistent  noncommutative quantum geometry, since 
the light cone function $\Delta (x^\nu)$ is not  a quantum operator and a classical metric has  been used for raising and lowering the indices.
  It projects the geometrical state onto a classical metric, and the operator labels onto a classical  inertial frame.  In the physical interpretation below,  the classical metric corresponds to a measurement, and thereby a choice of observer world line and  inertial frame. 

The  localization scale  $\tau_0$  has the same dimensions as $\hat{\tau}$. It fixes the information content of the system in physical units.  In our physical interpretation of this system,    $\Delta(x^\nu)$ is a  function of  physical space-time event positions, so $\tau_0$ represents a  quantization scale of light cone states in the  frame of an emergent observer--- a  finite resolution  in proper time.   As shown below, for a system that obeys the holographic principle, such as emergent  gravity,  $\tau_0= t_P.$  This  Planck-scale normalization will be assumed in the following.

\subsubsection{Nonlocal information, projection and uncertainty}

 Even though the degrees of freedom represented by the $\hat {\tau}_\kappa$'s have no local or dynamical effects,  their  fluctuations in time and direction affect correlations in  nonlocal measurements.

Consider  projection onto a 3D spacelike surface of constant $x^0\ne 0$.  The light cone function
$\Delta(x_0)$  is then a $\delta$-function on a 2-sphere of  radius $|x_0|$, which  coincides with the surface of a causal diamond.
For spatial positions  on this surface, we can set
 $\nu = 0$ in Eq. (\ref{qgeo})
to obtain a standard  spin algebra  in three dimensions,  with relabeled indices $i,j,k$ taking values $1,2,3$,
 \begin{equation}\label{3D}
[ \hat {\tau}_i, \hat {\tau}_j] = i \epsilon_{ijk}  \hat {\tau}_k  t_P.
\end{equation}
Thus, when a  light cone function $\Delta (x^\nu)$  is used to project the four dimensional space-time operators onto three dimensions --- an eigenstate of proper time--- it creates a fully consistent quantum  algebra that entangles curvature fluctuations in three dimensions.

We now recall some standard results of quantum mechanical spin in three dimensions (e.g., \cite{dicke,landau}). 
Positions in each spatial direction are related by
\begin{equation}\label{concrete3D}
[ \hat {\tau}_1, \hat {\tau}_2] =  i  \hat {\tau}_3  t_P,
\end{equation}
and its cyclic permutations.
A radial operator ${\hat {\tau}}^2\equiv \sum {\hat  {\tau}}_i^2$ commutes with all direction components:
\begin{equation}
[  \hat {\tau}_i, {\hat {\tau}}^2 ] =0.
\end{equation}
The discrete states of the system are assembled using  
 raising and lowering operators. For direction $3$,
\begin{equation}\label{raiselower}
{\hat {\tau}}_{3\pm}= {\hat {\tau}}_1 \pm i {\hat {\tau}}_2
\end{equation}
with the commuting properties
\begin{equation}
[{\hat {\tau}}_3, {\hat {\tau}}_{3\pm}] = \pm t_P {\hat {\tau}}_{3\pm},
\end{equation}
with raising and lowering operators for the other directions again obtained by cyclical permutations of $1,2,3$.
In a conventional integer-spin representation,   $  l= 0,1,3, \dots$ denotes the principal quantum number, and  $m = - l   \dots +l$ denotes projections onto a chosen axis, say $3$.  
The eigenvalues of ${\hat {\tau}}^2$  are $l (l+1) t_P^2$, and
the eigenvalues of ${\hat {\tau}}_3$ are $m_l t_P$.

  For each $l$ there are $2l+1$ directional projection eigenstates,
so the  number of degrees of freedom ${\cal N}$ scales holographically, as the  surface area in Planck units:  
\begin{equation}\label{eigenstates}
{\cal N}= \sum_{l'=0}^{l} (2l'+1)\approx  ( |{\tau}|/t_P)^2,
\end{equation}
where the approximation applies in the large $l$ limit.
Thus,  a Planck scale normalization  agrees with  
 holographic emergent gravity\cite{Jacobson1995,Verlinde2011,Padmanabhan:2013nxa,Jacobson:2015hqa}.  
We interpret ${\cal N}$  as the geometrical information in the whole (3+1D) volume enclosed by a  (2+1D) causal diamond defined by a 1D proper time interval.

Using standard algebraic methods\cite{PhysRevD.99.063531,dicke,landau}, 
it can be shown
that  in an eigenstate of ${\hat {\tau}}_3$,
\begin{equation}\label{orthogonal}
\langle {\hat {\tau}}_1\rangle=\langle {\hat {\tau}}_2\rangle = 0
\end{equation}
and
\begin{equation}\label{orthogonalvariance}
\langle {\hat {\tau}}_1^2\rangle=\langle {\hat {\tau}}_2^2\rangle = \langle{\hat {\tau}}^2-{\hat {\tau}}_3^2\rangle/2 =  t_P^2 [ l(l+1) - m^2]/2.
\end{equation}
Since $m^2\le l^2$, we can write a generalized uncertainty principle for  quantum fluctuations in any three orthogonal directions:
\begin{equation}\label{spinuncertainty}
\langle \delta {\tau}^2\rangle \equiv \langle \delta {\hat {\tau}}_1^2\rangle + \langle \delta {\hat {\tau}}_2^2\rangle +\langle \delta {\hat {\tau}}_3^2\rangle \ge   l \  t_P^2 > |{\tau}| t_P,
\end{equation}
where 
\begin{equation}
\delta {\hat {\tau}}_i^2 \equiv\langle{\hat {\tau}_i}^2\rangle-\langle{\hat {\tau}}_i\rangle^2.
\end{equation}
For large $l$, the  uncertainty (Eq. \ref{spinuncertainty}) is  much
 larger than $t_P^2$.
 In our interpretation of this system, 
 the $\delta {\hat {\tau}}_i$'s represent fluctuations in time as function of direction on  causal diamond surface of fixed radius, so the fractional fluctuations 
 \begin{equation}
\langle \delta {\tau}^2\rangle/ |{\tau}|^2 = t_P/|{\tau}|
\end{equation}
represent perturbations in gravitational redshift, dimensionless potential or curvature as a function of direction on a surface of radius $c|{\tau}|$.
  A causal diamond in flat space-time, or a horizon during inflation, has a 2D bounding surface radius with coherent  quantum fluctuations of this magnitude (as in Eq. \ref{potentialvariance}), nonlocally entangled among all three directions.

\end{document}